\documentclass[hyper,a4paper,notoc]{JHEP}
\pdfoutput=1
\usepackage{graphicx}
\usepackage{graphics}
\usepackage[mathscr]{eucal}
\usepackage{amssymb}
\usepackage{amsmath}
\usepackage{xspace}
\usepackage{listings}

\newcommand{\DRbar}{{\ensuremath{\overline{\mathrm{DR}}}}}
\newcommand{\SARAH}{{\tt SARAH}\xspace}
\newcommand{\SPheno}{{\tt SPheno}\xspace}
\newcommand{\GeV}{\text{GeV}\xspace}
\newcommand{\sign}{\text{sign}\xspace}

\newcommand{\BL}{{\ensuremath{B-L}}\xspace}
\newcommand{\UBL}{{\ensuremath{U(1)_{B-L}}}\xspace}
\newcommand{\BLSSM}{BLSSM\xspace}

\newcommand{\vevs}{\textit{vev}s\xspace}
\newcommand{\msugra}{mSUGRA\xspace}
\newcommand{\blino}{BLino\xspace}
\newcommand{\higgsino}{Higgsino\xspace}
\newcommand{\boringBM}{BLI\xspace}
\newcommand{\lightHiggsBM}{BLII\xspace}
\newcommand{\higgsinoBM}{BLIII\xspace}
\newcommand{\blinoBM}{BLIV\xspace}
\newcommand{\bileptinoBM}{BLV\xspace}
\newcommand{\EQ}{eq.\xspace}
\newcommand{\EQS}{eqs.\xspace}
\newcommand{\TAB}{Tab.\xspace}
\newcommand{\FIG}{Fig.\xspace}

\newcommand{\EG}{\textit{e.g}.\xspace}
\newcommand{\IE}{\textit{i.e}.\xspace}
\def\gsim{\raise0.3ex\hbox{$\;>$\kern-0.75em\raise-1.1ex\hbox{$\sim\;$}}}

\title{Mass spectrum of the minimal SUSY \BL model}

\preprint{BONN-TH-2011-17}

\author{
Ben O'Leary$^{1,a}$, Werner Porod$^{1,b}$,Florian Staub$^{1,2,c}$ \\
$^1$Institut f\"ur Theoretische Physik und Astrophysik, Universit\"at
W\"urzburg,\\
97074  W\"urzburg, Germany\\
$^2$Physikalisches Institut der Universit\"at Bonn, \\
53115 Bonn, Germany\\
$^a$Email: \email{ben.oleary@physik.uni-wuerzburg.de} \\
$^b$Email: \email{porod@physik.uni-wuerzburg.de} \\
$^c$Email: \email{fnstaub@th.physik.uni-bonn.de} \\
}
\abstract{The origin of $R$-parity in supersymmetric models
can be explained if \BL is part of the gauge group. 
We discuss the mass spectrum of the minimal $U(1)_Y \times \UBL$
model based on a GUT implementation using CMSSM-like boundary conditions.
Here we focus in particular on the Higgs and neutralino sectors 
in this class of models. While the neutralinos can have masses
as low as 100 GeV, we show that the requirement of being
consistent with existing bounds on the $Z'$ implies that in general
the sfermions have masses in the multi-TeV range. In the 
extended Higgs sector
we show the existence of a second light state which, however, will
be difficult to observe, while having at the same time a SM-like
Higgs in a mass range of 123-126 GeV.
Moreover, we  propose a set of benchmark scenarios for 
phenomenological studies. On the technical side we demonstrate
that gauge kinetic mixing effects can be quite important, affecting
in particular the Higgs and the neutralino sectors. Not only can they
shift the mass of the lightest neutralino by about 10 
per-cent but also they can change the nature of neutralinos and Higgs bosons
in a significant way.
 }

\begin{document}
\maketitle
\tableofcontents
\section{Introduction}
Models with an additional $U(1)_{\BL}$ gauge symmetry at the TeV 
scale have recently
received considerable attention. On one hand, they are among
the simplest extensions of Standard Model (SM) gauge group with observable
consequences at the LHC
 \cite{Emam:2007dy,Basso:2008iv,Basso:2010yz,Basso:2010si}. 
On the other,
this class of models can help to
understand the origin of $R$-parity and its possible spontaneous
violation in supersymmetric models 
\cite{Khalil:2007dr,Barger:2008wn,FileviezPerez:2010ek}, as
well as the mechanism of leptogenesis \cite{Pelto:2010vq,Babu:2009pi}.
It has been shown that a gauge sector containing $U(1)_Y \times \UBL$ 
can be a result of an \(E_8 \times E_8\) heterotic string theory 
(and hence M-theory) \cite{Ambroso:2009sc}. 
While most studies of supersymmetric variants have  so far focused on the
effects of the additional gauge group far below the GUT scale, the
questions arise of whether this group can be unified at the high scale with
the SM gauge group and what the phenomenological consequences are.
A renormalization group equation
(RGE) analysis of such a model, assuming the unification of the gauge
groups, has been performed in
\cite{FileviezPerez:2010ek,Ambroso:2009jd}. However, the effects of
possible mixing between the two Abelian groups have been neglected so
far:  it is well known that in models with several \(U(1)\) gauge
groups, kinetic mixing terms
\begin{equation}
\label{eq:offfieldstrength}
- \chi_{ab}  \hat{F}^{a, \mu \nu} \hat{F}^b_{\mu \nu}, \quad a \neq b
\end{equation}
between the field strength tensors are allowed by gauge and Lorentz
invariance \cite{Holdom:1985ag}, as $\hat{F}^{a, \mu \nu}$ and 
$\hat{F}^{b, \mu \nu}$ are gauge invariant quantities by themselves,
see \EG \cite{Babu:1997st}. Even if these terms are absent at tree 
level at a particular scale, they might be generated by RGE effects
\cite{delAguila:1988jz,delAguila:1987st}. 

The impact of gauge kinetic mixing in generic extensions of the 
standard model (SM) and the MSSM has been studied so far in several
aspects.
For instance, one can show that the dark matter of the universe 
can be charged with respect to an additional $U(1)$ but
neutral with respect to the SM gauge group. However, here one can show
that there is a residual SM gauge interaction of the dark matter particles
due to the gauge kinetic mixing.
The consequences for the relic density and the cross sections 
concerning direct as well indirect detection of dark matter 
have been analyzed \cite{Mambrini:2010dq,Chun:2010ve,Mambrini:2011dw}. 
It has been shown that
these cosmological bounds are sometimes more severe than the bounds 
from electroweak precision data if the dark matter candidate interacts
dominantly due to kinetic mixing. Moreover, the 
kinetic mixing in the context of
supersymmetric hidden sector dark matter has been considered 
in \cite{Andreas:2011in} and the LHC phenomenology of a nearly decoupled sector
only interacting with the visible sector due to 
kinetic mixing has been elaborated in \cite{Weihs:2011wp}.

In this work, we discuss the mass spectrum of the model presented in
\cite{Khalil:2007dr,FileviezPerez:2010ek}. This minimal \BL extension of the
Minimal Supersymmetric Standard Model (MSSM) has a \UBL gauge group
tensored to the SM gauge groups and two bileptonic chiral superfields which are
gauge singlets under SM gauge groups. In addition, three
right-handed neutrinos are needed to ensure that \UBL is anomaly-free,
which provide the necessary ingredients to explain neutrino data. We refer to
 this model as the \BLSSM.

The focus of this paper is on the mass spectrum of this model
and resulting phenomenological aspects assuming 
\msugra-like boundary conditions at the GUT
scale and unification of the \BL coupling with the SM couplings.
In particular we will demonstrate that gauge kinetic mixing
effects are particularly important in the Higgs and neutralino
sectors. These effects do not only change the masses of these
particles but have quite some impact on their nature, \EG
they induce tree-level mixing which would be absent if these effects
were to be neglected. We will show that new light Higgs states are
possible without being in conflict with current data while having
at the same time a SM-like Higgs in the range close to 120 GeV.
We will focus here on the case of R-parity conservation 
and discuss the case of broken R-parity violation in a subsequent paper
\cite{workinprep2}.

In the usual CMSSM with the MSSM particle content,
the lightest neutralino is mainly bino-like. We show that in our
model the nature of this particle can be quite different and identify
regions where it is either mainly a $SU(2)_L$-doublet \higgsino, a \UBL-gaugino
which we dub the \blino, or a fermionic partner of the \UBL-breaking scalar
 which we dub the bileptino, since we call the scalar the bilepton for reasons
 given below. In the next section we introduce
the model and focus in particular on aspects related to the
spectrum. In section \ref{sect:numerics} we present
our numerical results and provide benchmark points with distinct
features and in section \ref{sect:conclusions} we draw our conclusions.
In the appendices we collect supplementary formulas
for mass matrices, anomalous dimensions and $\beta$-functions at 
lowest order needed for the discussion of the main features in
section. The corresponding formulas including
higher order effects can be easily computed using the input files 
for \SARAH
given in appendix~\ref{app:modelfiles}.

\section{The Model}
In this section we present the particle content of the model
considered. An important aspect is the $U(1)$ gauge kinetic mixing which
is discussed in some detail as it leads to significant changes in 
the spectrum. Although we include
loop corrections for the numerical analysis when calculating the masses, we
restrict ourselves in this section to tree-level expressions, as this is
sufficient for discussing the main differences with respect to the MSSM.

\subsection{Particle content and superpotential}
The model consists of three generations of matter particles
including right-handed neutrinos
which can, for example, be embedded in $SO(10)$ 16-plets. Moreover,
below the GUT scale the usual MSSM Higgs doublets are present
as well as two fields $\eta$ and $\bar{\eta}$ responsible
for the breaking of the \UBL. Furthermore,  $\eta$ is
responsible for generating a Majorana mass term for the right-handed
neutrinos and thus  we interpret the \BL charge of this field as its
 lepton number, and likewise for $\bar{\eta}$, and call these fields
 bileptons since they carry twice the lepton number of (anti-)neutrinos.
We summarize the
quantum numbers of the chiral superfields
 with respect to $U(1)_Y \times SU(2)_L
\times SU(3)_C \times \UBL$  in Table~\ref{tab:cSF}.
\begin{table} 
\centering
\begin{tabular}{|c|c|c|c|c|c|} 
\hline \hline 
Superfield & Spin 0 & Spin \(\frac{1}{2}\) & Generations & \((U(1)_Y\otimes\,
SU(2)_L\otimes\, SU(3)_C\otimes\, \UBL)\) \\ 
\hline 
\(\hat{Q}\) & \(\tilde{Q}\) & \(Q\) & 3
 & \((\frac{1}{6},{\bf 2},{\bf 3},\frac{1}{6}) \) \\ 
\(\hat{D}\) & \(\tilde{d}^c\) & \(d^c\) & 3
 & \((\frac{1}{3},{\bf 1},{\bf \overline{3}},-\frac{1}{6}) \) \\ 
\(\hat{U}\) & \(\tilde{u}^c\) & \(u^c\) & 3
 & \((-\frac{2}{3},{\bf 1},{\bf \overline{3}},-\frac{1}{6}) \) \\ 
\(\hat{L}\) & \(\tilde{L}\) & \(L\) & 3
 & \((-\frac{1}{2},{\bf 2},{\bf 1},-\frac{1}{2}) \) \\ 
\(\hat{E}\) & \(\tilde{e}^c\) & \(e^c\) & 3
 & \((1,{\bf 1},{\bf 1},\frac{1}{2}) \) \\ 
\(\hat{\nu}\) & \(\tilde{\nu}^c\) & \(\nu^c\) & 3
 & \((0,{\bf 1},{\bf 1},\frac{1}{2}) \) \\ 
\(\hat{H}_d\) & \(H_d\) & \(\tilde{H}_d\) & 1
 & \((-\frac{1}{2},{\bf 2},{\bf 1},0) \) \\ 
\(\hat{H}_u\) & \(H_u\) & \(\tilde{H}_u\) & 1
 & \((\frac{1}{2},{\bf 2},{\bf 1},0) \) \\ 
\(\hat{\eta}\) & \(\eta\) & \(\tilde{\eta}\) & 1
 & \((0,{\bf 1},{\bf 1},-1) \) \\ 
\(\hat{\bar{\eta}}\) & \(\bar{\eta}\) & \(\tilde{\bar{\eta}}\) & 1
 & \((0,{\bf 1},{\bf 1},1) \) \\ 
\hline \hline
\end{tabular} 
\caption{Chiral superfields and their quantum numbers.}
\label{tab:cSF}
\end{table}

The superpotential is given by
\begin{align} 
\nonumber 
W = & \, Y^{ij}_u\,\hat{U}_i\,\hat{Q}_j\,\hat{H}_u\,
- Y_d^{ij} \,\hat{D}_i\,\hat{Q}_j\,\hat{H}_d\,
- Y^{ij}_e \,\hat{E}_i\,\hat{L}_j\,\hat{H}_d\,+\mu\,\hat{H}_u\,\hat{H}_d\, \\
 & \, \, 
+Y^{ij}_{\nu}\,\hat{L}_i\,\hat{H}_u\,\hat{\nu}_j\,- \mu' \,\hat{\eta}\,\hat{\bar{\eta}}\,
+Y^{ij}_x\,\hat{\nu}_i\,\hat{\eta}\,\hat{\nu}_j\,
\label{eq:superpot}
\end{align} 
and we have the additional soft SUSY-breaking terms:
\begin{align}
\nonumber \mathscr{L}_{SB} = & \mathscr{L}_{MSSM}
 - \lambda_{\tilde{B}} \lambda_{\tilde{B}'} {M}_{B B'}
 - \frac{1}{2} \lambda_{\tilde{B}'} \lambda_{\tilde{B}'} {M}_{B'}
 - m_{\eta}^2 |\eta|^2 - m_{\bar{\eta}}^2 |\bar{\eta}|^2
 - {m_{\nu,ij}^{2}} (\tilde{\nu}_i^c)^* \tilde{\nu}_j^c \\
& - \eta \bar{\eta} B_{\mu'} + T^{ij}_{\nu}  H_u \tilde{\nu}_i^c \tilde{L}_j
 + T^{ij}_{x} \eta \tilde{\nu}_i^c \tilde{\nu}_j^c 
\end{align}
$i,j$ are generation indices. Without loss of generality one can take $B_\mu$
and $B_{\mu'}$ to be real. The extended gauge group breaks to
$SU(3)_C \otimes U(1)_{em}$ as the Higgs fields and bileptons receive vacuum
expectation values (\vevs):
\begin{align} 
H_d^0 = & \, \frac{1}{\sqrt{2}} \left(\sigma_{d} + v_d  + i \phi_{d} \right),
\hspace{1cm}
H_u^0 = \, \frac{1}{\sqrt{2}} \left(\sigma_{u} + v_u  + i \phi_{u} \right)\\ 
\eta
= & \, \frac{1}{\sqrt{2}} \left(\sigma_\eta + v_{\eta} + i \phi_{\eta} \right),
\hspace{1cm}
\bar{\eta}
= \, \frac{1}{\sqrt{2}} \left(\sigma_{\bar{\eta}} + v_{\bar{\eta}}
 + i \phi_{\bar{\eta}} \right)
\end{align} 
We define $\tan\beta' = \frac{v_{\eta}}{v_{\bar{\eta}}}$ in analogy to
the ratio of the MSSM \vevs ($\tan\beta = \frac{v_u}{v_d}$).

\subsection{Gauge kinetic mixing}
\label{subsec:kineticmixing}

As already mentioned in the introduction, the presence of two Abelian
gauge groups in combination with the given particle content gives 
rise to a new
effect absent in the MSSM or other SUSY models with just one Abelian
gauge group: the gauge kinetic mixing. This can be seen most easily by
inspecting the matrix of the
anomalous dimension, which at one loop is given by
\begin{equation}
\gamma_{ab} = \frac{1}{16 \pi^2} \mbox{Tr}Q_a Q_b \,,
\end{equation}
where the indices $a$ and $b$ run over all $U(1)$ groups and
the trace runs over all fields charged under the corresponding
$U(1)$ group.

For our model we 
obtain
\begin{equation}
\gamma = \frac{1}{16 \pi^2} N \left( \begin{array}{cc} 11 & 4 \\
 4 & 6 \end{array} \right) N.
\end{equation}
and we see that there are sizable off-diagonal elements. $N$ contains the GUT
normalization  of the two Abelian gauge groups. We will take as in
ref.~\cite{FileviezPerez:2010ek}   \(\sqrt{\frac{3}{5}}\) for
\(U(1)_{Y}\) and \(\sqrt{\frac{3}{2}}\) for
\UBL, \IE $N=\text{diag}(\sqrt{\frac{3}{5}},\sqrt{\frac{3}{2}})$. 
Hence, we obtain finally
\begin{equation}
\label{eq:gammaMatrix}
 \gamma = \frac{1}{16 \pi^2}
  \left( \begin{array}{cc} \frac{33}{5} & 6 \sqrt{\frac{2}{5}} \\
                           6 \sqrt{\frac{2}{5}} & 9 \end{array} \right) .
\end{equation}
Therefore, even if at the GUT scale
 the $U(1)$ kinetic  mixing terms
are zero, they are induced via RGE evaluation at lower scales.
In practice it turns out that 
it is easier to work with  non-canonical covariant
derivatives instead of off-diagonal field-strength tensors such as in
\EQ~(\ref{eq:offfieldstrength}). 
However, both approaches are equivalent \cite{Fonseca:2011vn}. 
Hence in the following, we consider covariant derivatives of the form
\begin{equation}
\label{eq:kovariantDerivative}
 D_\mu  = \partial_\mu - i Q_{\phi}^{T} G  A 
\end{equation}
where \(Q_{\phi}\) is a vector containing the charges of the field $\phi$ with
respect to the two Abelian gauge groups, $G$ is the gauge coupling matrix
\begin{equation}
 G = \left( \begin{array}{cc} g_{YY} & g_{YB} \\
                              g_{BY} & g_{BB} \end{array} \right)
\end{equation}
and $A$ contains the gauge bosons $A = ( A^Y_\mu, A^B_\mu )^T$.

As long as the two Abelian gauge groups are unbroken, we have still
the freedom to perform a change of basis: $A=(A^Y_\mu, A^B_\mu)
\rightarrow A'=((A^Y_\mu)', (A^B_\mu)') = R A$ where $R$
is an orthogonal matrix. 
It is possible to absorb this rotation of the gauge fields completely
in the definition of the gauge couplings without the necessity of
changing the charges, which can easily be seen using
\EQ~(\ref{eq:kovariantDerivative})
\begin{equation} 
 Q_{\phi}^T G A = Q_{\phi}^T G (R^T R) A = Q_{\phi}^T (G R^T) A'
 = Q_{\phi}^T \tilde{G} A'
\end{equation}
This freedom can be used to choose a basis such that electroweak 
precision data can be accommodated in an easy way. A convenient
choice is the basis where \(g_{B Y}=0\) as in this basis
only the Higgs doublets contribute to the entries in the
gauge boson mass matrix of the $U(1)_Y\otimes SU(2)_L$ sector
and the impact of $\eta$ and $\bar{\eta}$ is only in the off-diagonal
elements as discussed in section \ref{subsec:gaugebosons}.
Therefore we choose
the following basis at the electroweak scale
\cite{Chankowski:2006jk}:
\begin{align}
\label{eq:gYYp}
 g'_{YY}
 = & \frac{g_{YY} g_{B B} - g_{Y B} g_{B Y}}{\sqrt{g_{B B}^2 + g_{B Y}^2}}
 = g_1  \\
 g'_{BB} = & \sqrt{g_{B B}^2 + g_{B Y}^2} = g_{BL} \\
 \label{eq:gtilde}
 g'_{Y B}
 = & \frac{g_{Y B} g_{B B} + g_{B Y} g_{YY}}{\sqrt{g_{B B}^2 + g_{B Y}^2}}
 = \tilde{g} \\
 g'_{B Y} = & 0
\label{eq:gBYp}
\end{align}
This also leads to our condition for finding the GUT scale in the numerical
analysis:
\begin{equation}
 g_2 \equiv
 \frac{g_{YY} g_{B B} - g_{Y B} g_{B Y}}{\sqrt{g_{B B}^2 + g_{B Y}^2}}
\end{equation}
This is equivalent to a rotation of the general $2\times 2$
gauge coupling matrix at each energy scale to the triangle form and
using $g_1 = g_2$ as the GUT condition. Neglecting threshold
corrections, this leads in the case of
kinetic mixing to exactly the same GUT scale as in the MSSM
\cite{Suematsu:1998wm}. 

Immediate interesting consequences of the gauge kinetic mixing arise
in various sectors of the model as discussed in the subsequent sections: 
(i) it induces mixing at tree level between the $H_u$, $H_d$
and $\eta$, $\bar{\eta}$; (ii) additional D-terms contribute 
to the mass matrices of the squarks and sleptons;
(iii) off-diagonal soft-SUSY breaking
terms for the gauginos are induced via RGE evolution  
\cite{Fonseca:2011vn, Braam:2011xh} with important 
consequences for the neutralino sector as  discussed in
section \ref{sec:neutralinos}, even if at some fixed scale $M_{ab}=0$
for $a\ne b$.

\subsection{Tadpole equations}
\label{subsec:tadpoles}

We find for the four minimization conditions at tree level
{\allowdisplaybreaks
\begin{align} 
\label{eq:vd}
t_d &= v_d \left(
m_{H_d}^2 +|\mu|^2 
+ \frac{1}{8} \left( g_{1}^{2} +g_{2}^{2} +\tilde{g}^{2} \right) 
  \left(v_{d}^{2}- v_{u}^{2}  \right)
+ \frac{1}{4} \tilde{g} g_{BL} 
  \left(v_{\eta}^{2} - v_{\bar{\eta}}^{2}\right) \right)
- v_u B_{\mu}=0\\ 
t_u &= v_u \left(
m_{H_u}^2 +|\mu|^2 
+ \frac{1}{8} \left( g_{1}^{2} +g_{2}^{2} +\tilde{g}^{2} \right) 
  \left(v_{u}^{2}- v_{d}^{2}  \right)
+ \frac{1}{4} \tilde{g} g_{BL} 
  \left(v_{\bar{\eta}}^{2} - v_{\eta}^{2}\right) \right)
- v_d B_{\mu}=0\\ 
t_{\eta} &= v_{\eta} 
\left( m_{\eta}^2 + |\mu'|^2 
+ \frac{1}{4} \tilde{g} g_{BL} \left(v_{d}^{2}- v_{u}^{2} \right)
+ \frac{1}{2} g_{BL}^{2} \left(v_{\eta}^{2} - v_{\bar{\eta}}^{2} \right) 
 \right)  - v_{\bar{\eta}} B_{\mu'}=0\\ 
t_{\bar{\eta}} &= v_{\bar{\eta}} 
\left( m_{\bar{\eta}}^2 + |\mu'|^2 
+ \frac{1}{4} \tilde{g} g_{BL} \left(v_{u}^{2}- v_{d}^{2} \right)
+ \frac{1}{2} g_{BL}^{2} \left(v_{\bar{\eta}}^{2} - v_{\eta}^{2} \right) 
 \right)  - v_{\eta} B_{\mu'}=0
\label{eq:vetabar}
\end{align} 
}
We solve them 
with respect to \(\mu, B_\mu, \mu'\) and \(B_{\mu'}\) as these
parameters do not enter any of the RGEs of the other parameters. 
Using $x^2=v_{\eta}^{2} + v_{\bar{\eta}}^{2}$ and 
$v^2=v_{d}^{2}+ v_{u}^{2}$ we obtain 
\begin{align}
\label{eq:tadmu}
 |\mu|^2 = & \frac{1}{8} \Big(\Big(2 \tilde{g} g_{BL} x^{2} \cos(2 {\beta'})
    -4 m_{H_d}^2  + 4 m_{H_u}^2 \Big)\sec(2 \beta)
    -4 \Big(m_{H_d}^2 + m_{H_u}^2\Big)
 - \Big(g_{1}^{2} + \tilde{g}^{2} + g_{2}^{2}\Big)v^{2} \Big)\\
 B_\mu =&-\frac{1}{8} \Big(-2 \tilde{g} g_{BL} x^{2} \cos(2 {\beta'})
    + 4 m_{H_d}^2  -4 m_{H_u}^2
  + \Big(g_{1}^{2} + \tilde{g}^{2} + g_{2}^{2}\Big)v^{2} \cos(2 \beta)
   \Big)\tan(2 \beta )     \\
 |\mu'|^2 =& \frac{1}{4} \Big(-2 \Big(g_{BL}^{2} x^{2}
  + m_{\eta}^2 + m_{\bar{\eta}}^2\Big) + \Big(2 m_{\eta}^2  -2 m_{\bar{\eta}}^2
  + \tilde{g} g_{BL} v^{2} \cos(2 \beta)   \Big)\sec(2 {\beta'})   \Big) \\
\label{eq:tadBmuP}
 B_{\mu'} =&  \frac{1}{4} \Big(-2 g_{BL}^{2} x^{2} \cos(2 {\beta'}) 
   + 2 m_{\eta}^2  -2 m_{\bar{\eta}}^2  + \tilde{g} g_{BL} v^{2} \cos(2 \beta)
   \Big)\tan(2 {\beta'} )
\end{align}
$M_Z' \simeq g_{BL} x$ as we will show in section \ref{subsec:gaugebosons}
 and, thus, we find an approximate
relation between $M_Z'$ and $\mu'$
\begin{equation}
\label{eq:tadpole_MZp}
 M^2_{Z'} \simeq 
 - 2 |\mu'|^2 + \frac{4 (m_{\bar{\eta}}^2 - m_{\eta}^2 \tan^2 \beta')
- v^2 \tilde{g} g_{BL} \cos\beta (1+\tan\beta') }{2 (\tan^2 \beta' - 1) }
\end{equation}
A closer inspection of the system shows that either $m_{\bar{\eta}}^2$ or
$m_{\eta}^2$ has to become negative to break $U(1)_{B-L}$. For both parameters,
gauge couplings enter the RGEs, increasing their values when evolving
from the GUT scale to the electroweak scale. The Yukawa couplings 
$Y_\nu$ and $Y_x$ as well as the trilinear couplings $T_\nu$ and $T_x$
lead to a decrease, but at the one-loop level they only affect the
RGE for $m^2_{\eta}$. 
However, neutrino data require
$|Y_{\nu,ij}|$ to be very small in this model and thus they can be neglected
for these considerations. 
Therefore,
$m_{\bar{\eta}}$ will always be positive whereas 
$m_{\eta}^2$ can become negative for sufficient large $Y_x$
 and $T_x$.
 
\begin{figure}[t]
\begin{minipage}{\linewidth}
 \includegraphics[width=0.48\linewidth]{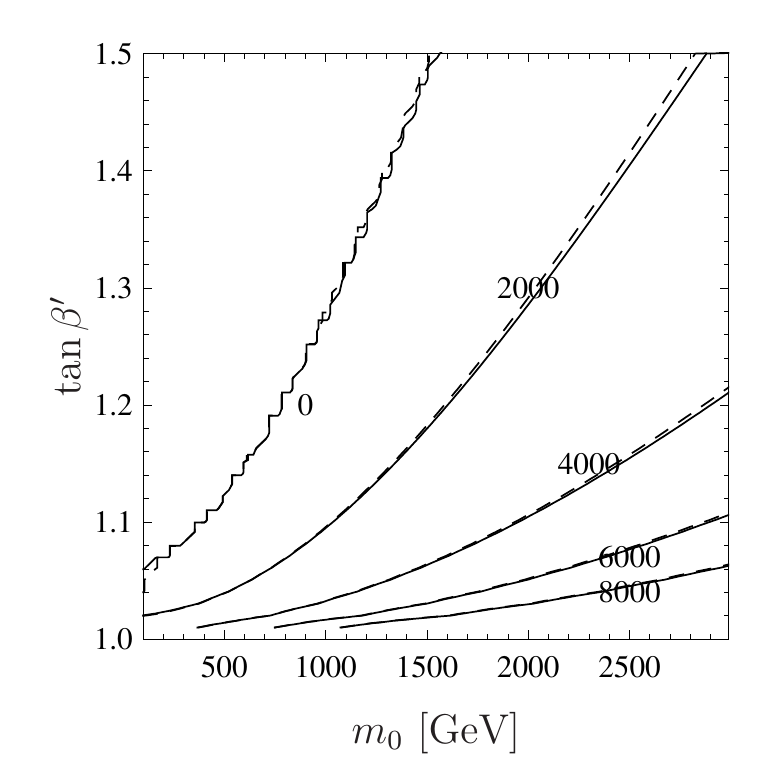}  
 \hfill 
 \includegraphics[width=0.48\linewidth]{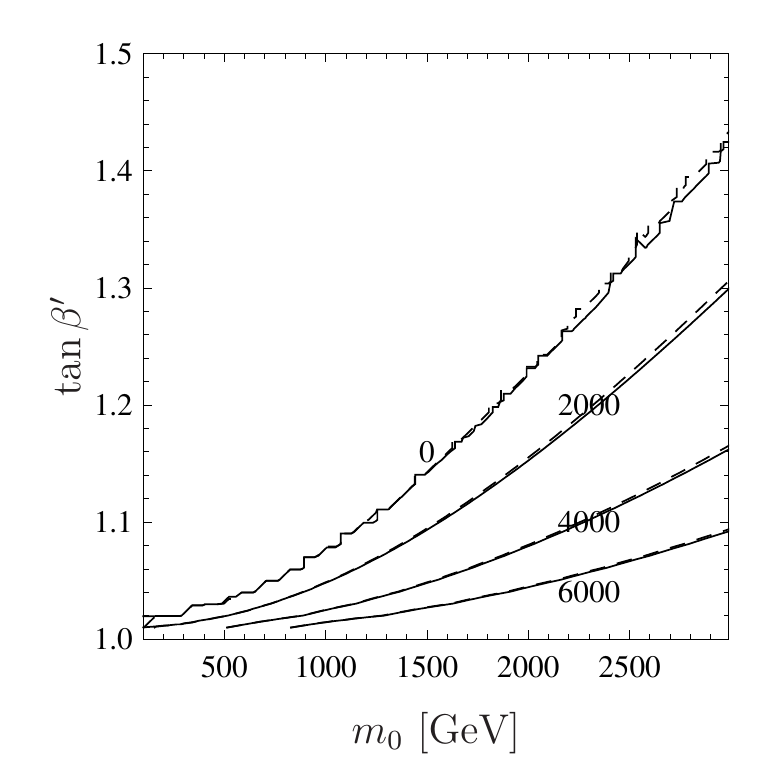}  
\end{minipage}
\caption{Contour plots of  $\mu'$ at tree-level in the $(m_0,\tan\beta')$-plane
for $M_{Z'}=2000~\GeV$ (left) and $M_{Z'}=4000~\GeV$ (right). The other
parameters are $M_{1/2} = 0.5$~TeV, $\tan(\beta)=10$, $A_0=1.5$~TeV, $Y_{x,ii}=0.42$. The
full lines correspond to the case including gauge kinetic mixing, the dashed
lines are  without kinetic mixing.}
\label{fig:MuP}
\end{figure}

 We can roughly estimate the contribution of these couplings to
 the running value of $m_{\eta}^2$ by a one-step integration 
assuming \msugra-like GUT conditions (see sect.~\ref{subsec:GUTconditions}) to
\begin{equation}
 \Delta m^2_{\eta} \simeq -\frac{1}{4\pi^2} \mbox{Tr}(Y_x Y^\dagger_x)
 (3 m^2_0 + A^2_0) \log\left(\frac{M_{GUT}}{M_{SUSY}}\right)
\end{equation}
with $T_x \simeq A_0 Y_x$. Therefore, we expect that large values of 
$m_0$ and $A_0$ will be
preferred, implying heavy sfermions.  Moreover,
$\tan\beta'$ has to be small and of $\mathscr{O}(1)$ in order to get a
small denominator in the second term of \EQ~\ref{eq:tadpole_MZp}.  One
last comment concerning the effect of gauge kinetic mixing: 
$g_{Y B}$ is
always negative below the GUT scale if it is zero at the GUT scale
as can be seen by the following: for vanishing
off-diagonal gauge couplings, the $\beta$-functions
\EQS~(\ref{eq:betaGBY}) and (\ref{eq:betaGYB}) will always be
positive, \IE $g_{B Y}$ and $g_{Y B}$ are driven negative. Using
\EQ~(\ref{eq:gtilde}), one can see that this also drives $\tilde{g}$
negative. Therefore, the second term will give a positive
contribution. From this point of view, one might expect that small
$m_0$ for given $\tan\beta'$ would be sufficient to get the same size of
$|\mu'|$. However, as can been seen in \FIG~\ref{fig:MuP}, where we
plot the tree-level value of $\mu'$ in the $(m_0, \tan\beta')$-plane
for the cases with and without kinetic mixing, the opposite effect
takes place. The reason is the contribution of the kinetic mixing to
the evaluation of $m_{\eta}^2$ and $m_{\bar{\eta}}^2$. One can also
see in this figure that the upper limit of $\tan\beta'$ for a given
value of $m_0$ decreases with increasing $M_{Z'}$ as expected. 
Even if one might get the impression from this figure that the 
effects of kinetic mixing are in general small as they slightly shift the region 
where breaking of \UBL can occur, it will be shown later
that it can have a significant impact on the masses.

For the numerical results we include one-loop corrections
to \EQS~(\ref{eq:vd})-(\ref{eq:tadBmuP}) as well as for all
masses. This is done by using  the \DRbar\
scheme and extending the MSSM results
given in ref.~\cite{Pierce:1996zz} in a similar manner to the NMSSM case
discussed in ref.~\cite{Staub:2010ty}. We denote the one-loop contributions to the
tadpole equations~(\ref{eq:vd})-(\ref{eq:vetabar}) by
\(\delta t^{(1)}_i\). The requirement of keeping the values
of $\tan\beta$ and $\tan\beta'$ after including the loop 
corrections as well as the conditions
\begin{equation}
 t_i + \delta t^{(1)}_i = 0 \qquad{\rm for}\quad i=d,u,\eta, \bar{\eta}
\label{eq:oneloop}
\end{equation}
lead to shifts of $\mu$, $B_\mu$, $\mu'$ and $B_{\mu'}$ compared to
the values obtained by \EQS~(\ref{eq:tadmu})-(\ref{eq:tadBmuP}).

\subsection{Gauge boson mixing}
\label{subsec:gaugebosons}
Due to the presence of the kinetic mixing terms, the $B'$ boson mixes
at tree level with the $B$ and $W^3$ bosons. Requiring the conditions of
\EQS~(\ref{eq:gYYp})-(\ref{eq:gBYp}) means that the corresponding mass
matrix reads, in the basis $(B,W^3,B')$,
\begin{align}
\label{eq:MMgauge}
\left(\begin{array}{ccc} 
\frac{1}{4} g_{1}^{2} v^2 & -\frac{1}{4} g_1 g_2 v^2
 & \frac{1}{4} g_1 \tilde{g} v^2\\
-\frac{1}{4} g_1 g_2 v^2 & \frac{1}{4} g_{2}^{2} v^2
 & -\frac{1}{4}\tilde{g} g_2 v^2\\
\frac{1}{4} g_1 \tilde{g} v^2 & -\frac{1}{4}\tilde{g} g_2 v^2
 & ( g_{BL}^{2} x^2 + \frac{1}{4} \tilde{g}^{2} v^2 )
\end{array} \right)
\end{align}
In the limit $\tilde{g} \rightarrow 0 $ both sectors decouple and the
upper $2\times 2$ block is just the standard mass matrix of the
neutral gauge bosons in EWSB. This mass matrix  can
be diagonalized by a unitary mixing matrix to get the physical mass
eigenstates $\gamma$, $Z$ and $Z'$. The rotation matrix can be expressed
by two mixing angles $\Theta_W$ and $\Theta'_W$ as
\begin{align} 
\left(\begin{array}{c} 
B\\ 
W\\ 
{B'}\end{array} \right) 
 = & \,\left( 
\begin{array}{ccc} 
\cos\Theta_W & \cos{\Theta'}_W \sin\Theta_W & - \sin\Theta_W \sin{\Theta'}_W \\ 
\sin\Theta_W & - \cos\Theta_W \cos{\Theta'}_W & \cos\Theta_W \sin{\Theta'}_W \\ 
 0 & \sin{\Theta'}_W  & \cos{\Theta'}_W \end{array} 
\right)
\left(\begin{array}{c} 
\gamma\\ 
Z\\ 
{Z'}\end{array} \right) 
\end{align}
The third angle is zero due to the special form of this matrix. ${\Theta'}_W$
 can be approximated by \cite{arXiv:1004.3039}
\begin{equation}
\label{eq:ThetaWP}
\tan 2 {\Theta'}_W \simeq
 \frac{2 \tilde{g} \sqrt{g_1^2 + g_2^2}}{\tilde{g}^2
 + 16 \left(\frac{x}{v}\right)^2 g_{BL}^2 -g_2^2 - g_1^2}
\end{equation}
The exact eigenvalues of \EQ~(\ref{eq:MMgauge}) are given by
\begin{align}
M_{\gamma} &= 0 \\
M_{Z,Z'} & = \frac{1}{8}\Big((g_1^2+g_2^2+\tilde{g}^2)v^2
 + 4 g_{BL}^2 x^2 \mp \nonumber \\
 & \hspace{1.5cm} \sqrt{(g_1^2+g_2^2+\tilde{g}^2)^2 v^4
 - 8 (g_1^2+g_2^2)g_{BL}^2 v^2 x^2 + 16 g_{BL}^2 x^4}\Big)
\end{align}
Expanding these formulas in powers of $v^2/x^2$, we find up to first
order:
\begin{equation}
\label{eq:MassZP}
M_Z = \frac{1}{4}\left(g_1^2 + g_2^2\right) v^2 \,,\hspace{1cm}
 M_{Z'} = g_{BL}^2 x^2 + \frac{1}{4} \tilde{g}^2 v^2
\end{equation}
All parameters in \EQS~(\ref{eq:vd})-(\ref{eq:vetabar}) as well as in the
 following mass matrices
 are understood as
running parameters at a given renormalization scale $Q$. Note that the \vevs
 $v_d$ and $v_u$
are obtained from the running mass $M_Z(Q)$ of the $Z$ boson, 
which is related to the pole mass $M_Z$ through
\begin{equation}
M^2_Z(Q) = \frac{g^2_1+g^2_2}{4} (v^2_u+v^2_d)
 = M^2_Z + \mathrm{Re}\big\{ \Pi^T_{ZZ}(M^2_Z) \big\}.  
\end{equation}
Here, $\Pi^T_{ZZ}$ is the transverse self-energy of the $Z$. See for more
 details also ref.~\cite{Pierce:1996zz}.

The mass of additional vector bosons as well as their mixing with
the SM $Z$ boson, which imply for example a deviation of the fermion
couplings to the $Z$ boson compared to SM expectations, is severely
constrainted by precision measurements from the LEP experiments
\cite{Alcaraz:2006mx,Erler:2009jh,Nakamura:2010zzi}. 
The bounds are on both the mass of the $Z'$ and the mixing with
the standard $Z$ boson, where the latter is constrained by
$|\sin(\Theta_{W'})<0.0002|$. Using \EQ~(\ref{eq:ThetaWP}) together with
 \EQ~(\ref{eq:MassZP}) as well as the values of the running gauge couplings,
 a limit on the $Z'$ mass of about 1.2~TeV is obtained. Taking in addition the
bounds obtained from $U$, $T$ and $S$ parameters into account
 \cite{Cacciapaglia:2006pk} one gets $\frac{M_{Z'}}{Q_e^{B-L} g_{B-L}}>7.1$~TeV
 which for $g_{B_L}\simeq 0.52$ implies $M_{Z'} \gsim 1.8$~TeV. Therefore we
 have taken always $M_{Z'} \geq 2$~TeV. In this way we have also satisfied the
 most recent bounds obtained by the ATLAS and CMS \cite{Adams:Moriond12}.

\subsection{The Higgs sector}
\label{sec:model_higgs}
In this section we present the tree-level formulas for the
Higgs sector and we briefly discuss the main steps to include
 the one-loop corrections. The one-loop formulas and further
 details will be presented elsewhere \cite{workinprep}.
 
\subsubsection{Pseudoscalar Higgs bosons}
It turns out that in this sector there is no mixing between 
the $SU(2)$ doublets and the
bileptons at tree level and we obtain in the basis 
$(\phi_d,\phi_u,\phi_\eta,\phi_{\bar{\eta}})$:
\begin{equation}
m^2_{A,T} = \left(\begin{array}{cccc}
B_\mu \tan\beta & B_\mu & 0 & 0 \\
B_\mu  & B_\mu \cot\beta & 0 & 0 \\
 0 & 0  &B_{\mu'} \tan\beta' & B_{\mu'} \\
 0 & 0  &B_{\mu'} & B_{\mu'} \cot\beta' 
\end{array}
\right) \,.
\label{eq:mA2}
\end{equation}
Obviously, both sectors decouple at tree level. This is a consequence of the
 fact that we assume that there is no CP violation in the Higgs sector, so the
 different D-term contributions cancel exactly. 
One obtains two physical states $A^0$ and $A^0_\eta$ with masses
\begin{equation}
m^2_{A^0} = \frac{2 B_\mu}{\sin2\beta} \thickspace, \hspace{1cm} 
m^2_{A^0_\eta}  = \frac{2 B_{\mu'}}{\sin2\beta'} \thickspace.
\end{equation}
A more detailed study of the pseudoscalar sector at one-loop, 
including the question if the block-diagonal form of the mass 
matrix in \EQ~(\ref{eq:mA2}) can be maintained at higher order,
goes beyond the scope of this work and will be presented elsewhere \cite{workinprep}.

\subsubsection{Scalar Higgs bosons}
 \label{sec:subsubScalars}
 
In the scalar sector the gauge kinetic terms do induce a mixing
between the $SU(2)$ doublet Higgs fields and the bileptons.
The mass matrix reads at tree level in the basis 
$(\sigma_d,\sigma_u,\sigma_\eta,\sigma_{\bar{\eta}})$:
\begin{align}
& m^2_{h,T} = \nonumber \\
&\left(\begin{array}{cccc}
m^2_{A^0} s^2_\beta + \bar{g}^2 v^2_u & \,\,
-m^2_{A^0} c_\beta s_\beta - \bar{g}^2 v_d v_u &
 \frac{\tilde{g} g_{BL}}{2}   v_d v_{\eta} &
  -\frac{\tilde{g} g_{BL}}{2} v_d v_{\bar{\eta}} \\
-m^2_{A^0} c_\beta s_\beta  - \bar{g}^2 v_d v_u &
m^2_{A^0} c^2_\beta + \bar{g}^2 v^2_d & \,\,
 - \frac{\tilde{g} g_{BL}}{2}   v_u v_{\eta} &
  \frac{\tilde{g} g_{BL}}{2} v_u v_{\bar{\eta}} \\
\frac{\tilde{g} g_{BL}}{2}   v_d v_{\eta} &
 - \frac{\tilde{g} g_{BL}}{2}   v_u v_{\eta}  &
  m^2_{A^0_\eta} c^2_{\beta'} + g^2_{BL} v^2_\eta &
 \,\, -  m^2_{A^0_\eta} c_{\beta'} s_{\beta'} 
  - g^2_{BL} v_\eta v_{\bar{\eta}} \\
-\frac{\tilde{g} g_{BL}}{2} v_d v_{\bar{\eta}} &
 \frac{\tilde{g} g_{BL}}{2} v_u v_{\bar{\eta}}  &
\,\, -  m^2_{A^0_\eta} c_{\beta'} s_{\beta'} 
  - g^2_{BL} v_\eta v_{\bar{\eta}} &
 m^2_{A^0_\eta} s^2_{\beta'} + g^2_{BL} v^2_{\bar{\eta}}
\end{array}
\right)
\end{align}
where we have defined $\bar{g}^2 = \frac{1}{4} (g^2_1+g^2_2+\tilde g^2)$,
$c_x = \cos(x)$ and $s_x = \sin(x)$ ($x=\beta,\beta')$.
The one-loop corrections  are included by calculating
 the real part of the poles of the
corresponding propagator matrices \cite{Pierce:1996zz,workinprep}
\begin{equation}
\mathrm{Det}\left[ p^2_i \mathbf{1} - m^2_{h,1L}(p^2) \right] = 0,
\label{eq:propagator}
\end{equation}
where
\begin{equation}
 m^2_{h,1L}(p^2) = m^{2,h}_T -  \Pi_{hh}(p^2) .
\end{equation}
Equation (\ref{eq:propagator}) has to be solved for each
eigenvalue $p^2=m^2_i$ which can be achieved in an iterative
procedure.

\subsubsection{The charged Higgs boson}

At the tree level one finds that the charged Higgs boson mass has
exactly the same form as in the MSSM:
\begin{equation}
m^2_{H^+} = m^2_{A^0} + m^2_W
\end{equation}
However, for the one-loop corrections one obtains additional
contributions due to the kinetic gauge mixing \cite{workinprep}.

\subsection{Neutralinos}

In the neutralino sector we find that the gauge kinetic effects lead to a mixing
 between the usual MSSM neutralinos with the additional states, similar to the
 mixing in the CP-even Higgs sector. In other words, were these to be neglected,
 both sectors would decouple.
 The mass matrix reads in
the basis \( \left(\lambda_{\tilde{B}}, \tilde{W}^0, \tilde{H}_d^0,
  \tilde{H}_u^0, \lambda_{\tilde{B}{}'}, \tilde{\eta},
  \tilde{\bar{\eta}}\right) \)
\begin{equation} 
\label{eq:NeutralinoMM}
m_{\tilde{\chi}^0} = \left( 
\begin{array}{ccccccc}
M_1 & 0 & -\frac{1}{2} g_1 v_d & \frac{1}{2} g_1 v_u & \frac{1}{2} {M}_{B B'}
 & 0 & 0 \\ 
0 & M_2 & \frac{1}{2} g_2 v_d  & -\frac{1}{2} g_2 v_u  & 0 & 0 & 0 \\ 
-\frac{1}{2} g_1 v_d  & \frac{1}{2} g_2 v_d  & 0 & - \mu
 & -\frac{1}{2} \tilde{g} v_d  & 0 & 0 \\ 
\frac{1}{2} g_1 v_u  & -\frac{1}{2} g_2 v_u & - \mu  & 0
 & \frac{1}{2} \tilde{g} v_u  & 0 & 0 \\ 
\frac{1}{2} {M}_{B B'}  & 0 & -\frac{1}{2} \tilde{g} v_d
 & \frac{1}{2} \tilde{g} v_u  & {M}_{B} & - g_{BL} v_{\eta}
 & g_{BL} v_{\bar{\eta}} \\ 
0 & 0 & 0 & 0 & - g_{BL} v_{\eta}  & 0 & - {\mu'} \\ 
0 & 0 & 0 & 0 & g_{BL} v_{\bar{\eta}}  & - {\mu'} & 0\end{array} 
\right) 
\end{equation} 
It is well known that for real parameters such a matrix can be diagonalized
by an orthogonal mixing matrix $N$ such that
$N^* M^{\tilde\chi^0}_T N^\dagger$ is diagonal. For complex parameters
one has to diagonalize $M^{\tilde\chi^0}_T (M^{\tilde\chi^0}_T)^\dagger$. 
We obtain, in a straightforward generalization
of the formulas given in \cite{Pierce:1996zz},
at the one-loop level 
\begin{eqnarray}
M^{\tilde\chi^0}_{1L} (p^2_i) &=& M^{\tilde\chi^0}_T - 
\frac{1}{2} \bigg[ \Sigma^0_S(p^2_i) + \Sigma^{0,T}_S(p^2_i)
 + \left(\Sigma^{0,T}_L(p^2_i)+   \Sigma^0_R(p^2_i)\right) M^{\tilde\chi^0}_T
 \nonumber \\
&& \hspace{16mm}
+ M^{\tilde\chi^0}_T \left(\Sigma^{0,T}_R(p^2_i) +  \Sigma^0_L(p^2_i) \right)
 \bigg] ,
\end{eqnarray}
where we have denoted the wave-function corrections by $\Sigma^{0}_R$,
 $\Sigma^{0}_L$ and the direct one-loop contribution to the mass by 
$\Sigma^{0}_S$. 

In this model, for the chosen boundary conditions, the lightest
 supersymmetric particle (LSP), and therefore the dark matter candidate, is 
always either the lightest neutralino or the lightest sneutrino. The reason is
that $m_0$ must be very heavy in order to solve the tadpole equations,
and therefore all sfermions are heavier than the lightest neutralino, with the
possible exception of the sneutrinos. As described in
 sec.~\ref{subsec:charginosAndSfermions}
 below, the splitting of the CP-even and CP-odd components of the sneutrinos can
 be very large, pushing the mass of the lighter eigenstate down even to the
 point of being lighter than the lightest neutralino. However, we
 have chosen to leave the investigation of such a scenario to future work, and
 we take only points with neutralino LSPs as our benchmark scenarios. Before
 leaving this topic, we note that \bileptinoBM has a lightest sneutrino that is
 almost as light as the lightest neutralino.
 A neutralino LSP is in general a mixture of all seven gauge eigenstates.
However, normally the character is dominated by only one or two
constituents. In that context, we can distinguish the following extreme
cases:
\begin{enumerate}
 \item $M_1 \ll M_2, \mu, M_B, \mu'$: Bino-like LSP
 \item $M_2 \ll M_1, \mu, M_B, \mu'$: Wino-like LSP
 \item $\mu \ll M_1, M_2, M_B, \mu'$: Higgsino-like LSP
 \item $M_B \ll M_1, M_2, \mu, \mu'$: \blino-like LSP
 \item $\mu' \ll M_1,M_2,\mu, M_B$: Bileptino-like LSP
\end{enumerate}
In addition, we will summarize the Bino- and Wino-like states, \textit{i.e.}
 the states built by the gauginos of the MSSM, in the 
following often by `gaugino-like'. Note that this doesn't include the \blino, the
 gaugino of the \BL sector. 

Although the gauge kinetic effects do lead to sizable effects in
the spectrum, they are not large enough to lead to a large mixing
between the usual MSSM-like states and the new ones. Therefore,
we find that the LSP is either mainly a MSSM-like state or mainly an
admixture between the \blino and the bileptinos.
 A discussion of the
parameter space where the different characters appear is given in
sec.~\ref{sec:neutralinos}.

\subsection{Charginos and sfermions}
\label{subsec:charginosAndSfermions}

For completeness we also give a short summary of the other sectors of the
model. The chargino mass matrix at tree level is exactly the same as
for the MSSM:
\begin{equation}
M^{\tilde\chi^+}_T = \left( 
\begin{array}{cc}
M_2 &\frac{1}{\sqrt{2}} g_2 v_u \\ 
\frac{1}{\sqrt{2}} g_2 v_d  & \mu \end{array}  
\right) .
\end{equation} 
This mass matrix is diagonalized by a biunitary transformation such that
$U^* M^{\tilde\chi^+}_T V^\dagger$ is diagonal. The matrices $U$ and $V$
are obtained by diagonalizing
 $ M^{\tilde\chi^+}_T ( M^{\tilde\chi^+}_T)^\dagger$
and $(M^{\tilde\chi^+}_T)^* ( M^{\tilde\chi^+}_T)^T$, respectively. At the
 one-loop level, one has to add the self-energies \cite{Pierce:1996zz}
\begin{eqnarray}
M^{\tilde\chi^+}_{1L}(p^2_i) =  M^{\tilde\chi^+}_T - \Sigma^+_S(p^2_i)
 - \Sigma^+_R(p^2_i) M^{\tilde \chi^+}_T
 - M^{\tilde \chi^+}_T \Sigma^+_L(p^2_i) .
\end{eqnarray}

The mass matrices for the squarks and the charged sleptons are given in
appendix~\ref{app:massmatrices}. At tree level, the differences compared
to the MSSM are the additional D-terms in the diagonal
entries.  All complex scalar mass matrices are diagonalized by an unitary matrix $Z$ 
 \begin{equation}
 Z_\phi m^2_\phi Z^{\dagger}_\phi = m^{2}_{\phi,\mathrm{diag}} .
 \end{equation}
 The corresponding mass matrices at the one-loop level are again obtained 
 by taking into account the self-energy according to
 \begin{equation}
 	m^{2,\phi}_{1L}(p^2_i) = m^{2,\phi}_{T} - \Pi_{\phi \phi}(p^2_i) ,
 \end{equation}
 and the one-loop masses are obtained by calculating the poles
 of the real part of the propagator matrix. 

We focus here on the sneutrino sector as it shows two distinct features
compared to the MSSM. Firstly, it gets enlarged by the
additional superpartners of the right-handed neutrinos. Secondly,
even more drastically, a splitting between the real and imaginary
parts of the sneutrino occurs resulting in twelve states: six scalar
sneutrinos and six pseudoscalar ones
\cite{Hirsch:1997vz,Grossman:1997is}. The origin of this splitting
is the $Y^{ij}_x\,\hat{\nu}_i\,\hat{\eta}\,\hat{\nu}_j$ in the
superpotential, eq.~(\ref{eq:superpot}), which is a $\Delta L=2$ operator.  
Therefore, we define
\begin{equation}
\tilde{\nu}^i_L = \frac{1}{\sqrt{2}}\left(\sigma^i_L + i \phi^i_L\right)\, \hspace{1cm} 
 \tilde{\nu}^i_R = \frac{1}{\sqrt{2}}\left(\sigma^i_R + i \phi^i_R\right)
\end{equation}
The $6 \times 6$ mass matrices of the CP-even and CP-odd
sneutrinos can be written in the basis
 \(\left(\sigma_{L},\sigma_{R}\right)\) respectively  \(\left(\phi_{L},\phi_{R}\right)\) as
\begin{equation} 
m^2_{\nu^{R,I}} = \Re\left( 
\begin{array}{cc}
m_{LL} &m^T_{RL}\\ 
m_{RL} &m_{RR}\end{array} 
\right) 
\end{equation} 
with
\begin{align} 
m_{LL} &= \frac{1}{8} \Big({\bf 1} \Big( \Big(g_{1}^{2}+g_{2}^{2}+\tilde{g}^{2} \Big) \Big(- v_{u}^{2}  + v_{d}^{2}\Big) +\tilde{g} g_{B} \Big(-2 v_{\bar{\eta}}^{2}  + 2 v_{\eta}^{2}  - v_{u}^{2}  + v_{d}^{2}\Big)\nonumber \\ 
 & \, \, +2 g_{B}^{2} \Big(- v_{\bar{\eta}}^{2}  + v_{\eta}^{2}\Big)\Big)+8 m_l^2 +4 v_{u}^{2} {Y_\nu  Y_{\nu}^{\dagger}} \Big)\\ 
m_{RL} &= \frac{1}{4} \Big(-2 \sqrt{2} v_d \mu Y_{\nu}^{\dagger}   + v_u \Big(2 \sqrt{2} T_{\nu}^{T}  \pm 4 v_{\eta} {Y_x  Y_{\nu}^{\dagger}} \Big)\Big)\\ 
\label{eq:mRR}
m_{RR} &= \frac{1}{8} \Big({\bf 1}\Big( 2 g_{B}^{2} (v^2_{\bar{\eta}}  - v^2_{\eta}) - \tilde{g} g_{B} \Big(- v_{u}^{2}  + v_{d}^{2}\Big)\Big)+8 {m_\nu^2} +2 v_{\bar{\eta}} \Big(\mp 4 \sqrt{2} Y_x {\mu'}^*  \Big) \nonumber \\ 
 & \, \, +4 v_{u}^{2} {Y_{\nu}^{T}  Y_\nu^*} 
 +2 v_{\eta} \Big(\pm 4 \sqrt{2} T_x  + 8 v_{\eta}  {Y_x  Y_x^*}\Big)\Big)
\end{align} 
where we have assumed CP conservation. In the case of complex trilinear
couplings, a mixing between the scalar and pseudoscalar particles occurs,
resulting in 12 mixed states and consequently in a $12\times 12$ mass 
matrix.
It particular the term $\sim v_{\bar{\eta}} Y_x {\mu'}^*$ is potentially
large and induces a large mass splitting between the scalar and
pseudoscalar states. Also the corresponding soft SUSY term
$\sim v_{\eta} T_x$ can lead to a sizable mass splitting in the
case of large $|A_0|$. As a side-remark we note that in such a 
case also the determinant of this matrix could become negative
indicating the breaking of R-parity in a somewhat different way compared to 
the discussion in ref.~\cite{FileviezPerez:2010ek}. However, here we will
 concentrate on the R-parity conserving case and discuss R-parity 
violation elsewhere \cite{workinprep2}.

\subsection{Boundary conditions at the GUT scale}
\label{subsec:GUTconditions}

We will consider in the following a scenario motivated by minimal
supergravity. This means that we assume a GUT unification of all
soft-breaking scalar masses as well as a unification of all gaugino mass
 parameters
\begin{align}
 m^2_0 = & m^2_{H_d} = m^2_{H_u} = m^2_{\eta} = m^2_{\bar{\eta}} \\
m^2_0\delta_{ij} = & m_D^2 \delta_{ij} =  m_U^2 \delta_{ij} = m_Q^2 \delta_{ij}
= m_E^2 \delta_{ij} = m_L^2 \delta_{ij} = m_{\nu}^2 \delta_{ij} \\
 M_{1/2} = & M_1 = M_2 = M_3 = M_{\tilde{B}'}
\end{align}
Also, for the trilinear soft-breaking coupling, the ordinary \msugra conditions
 are assumed
\begin{align}
 T_i = A_0 Y_i, \hspace{1cm} i = e,d,u,x,\nu \thickspace . 
\end{align}
We do not fix the parameters \(\mu, B_\mu, \mu'\) and \(B_{\mu'}\) 
at the GUT scale but determine them from the tadpole equations.
The reason is that they do not enter the RGEs of the other
parameters and thus can be treated independently. The
corresponding formulas are given in  section \ref{subsec:tadpoles}.

In addition, we consider the mass of the $Z'$ and $\tan\beta'$ as 
inputs and use the following set of free parameters
\begin{eqnarray}
& m_0, \thickspace M_{1/2},\thickspace A_0,\thickspace \tan\beta,\thickspace
 \tan\beta',\thickspace \sign(\mu),\thickspace \sign(\mu'),\thickspace M_{Z'},
 \thickspace  Y_x \thickspace \mbox{and} \thickspace Y_{\nu} . &
\end{eqnarray}
\(Y_{\nu}\) is constrained by neutrino data and must therefore be
very small in comparison to the other couplings; thus it
can be neglected in the following. $Y_x$ can always be taken
diagonal and thus effectively we have 9 free parameters and two signs. 

\begin{figure}[t]
\includegraphics[width=0.48\linewidth]{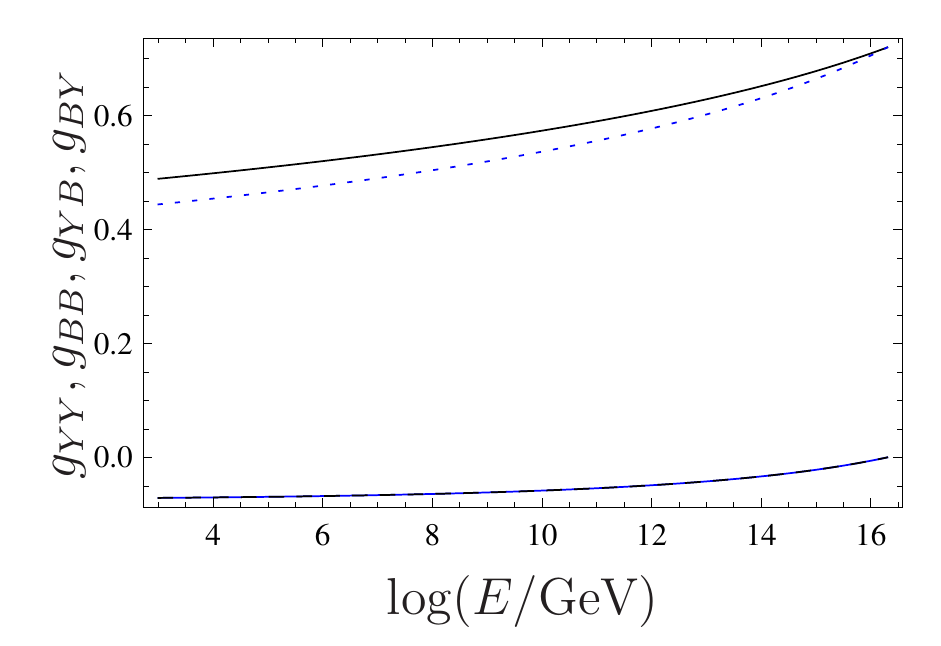}  
\hfill
\includegraphics[width=0.48\linewidth]{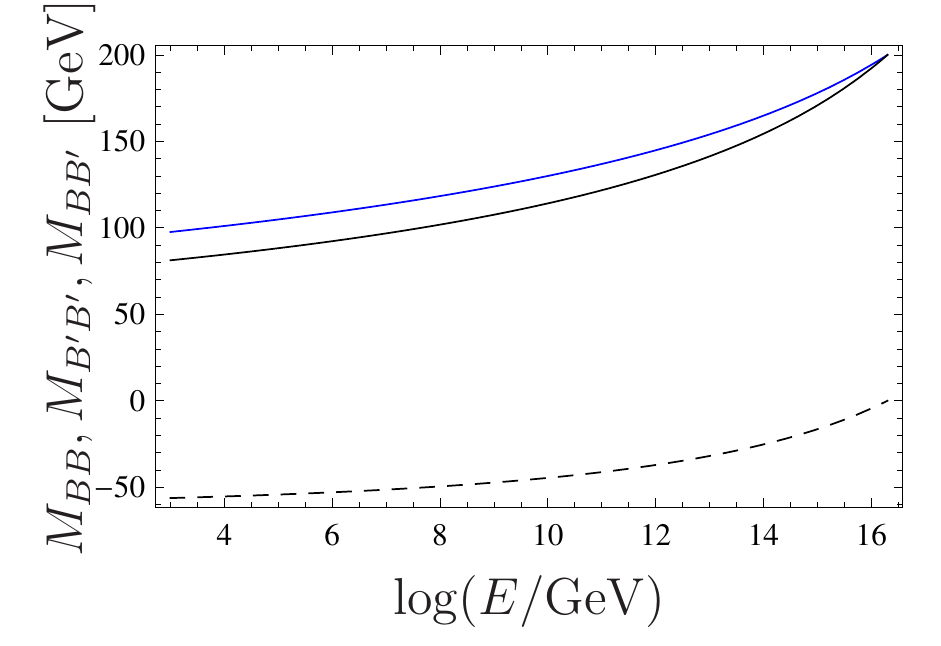}  
\caption{One-loop evaluation of the gauge couplings (left) and gaugino mass
  parameters (right) associated with the Abelian gauge groups. Blue color is
 used for $g_{YY}$ \& $M_{BB}$, black
  for $g_{BB}$ \& $M_{B'B'}$, dashed black for $g_{YB}$ \& $M_{BB'}$ and dashed blue for
  $g_{BY}$. The evolution of the two off-diagonal couplings hardly differs
  and thus the two lines nearly match.}
\label{fig:running}
\end{figure}

Furthermore, we assume that there are no off-diagonal gauge couplings
or gaugino mass parameters present at the GUT scale
\begin{align}
 g_{B Y} = & g_{Y B} = 0 \\
 M_{B B'} = & 0
\end{align}
This choice is motivated by the possibility that the two Abelian groups are a
remnant of a larger product group which gets broken at the GUT
scale as stated in the introduction.
In that case \(g_{YY}\) and \(g_{B B}\) correspond to the physical
couplings $g_1$ and $g_{BL}$, for which we assume a unification with
$g_2$:
\begin{equation}
 g^{GUT}_1 = g^{GUT}_2 = g_{BL} \thickspace .
\end{equation}
where we have already taken into account the correct GUT normalization
as discussed in section \ref{subsec:kineticmixing}.
In \FIG~\ref{fig:running} we display the running of the 
 gauge couplings and gaugino parameters in the Abelian sector
to demonstrate the effect of the gauge
kinetic mixing.  The GUT scale has been
set to $2\cdot 10^{16}$~GeV where we have fixed 
 $g_{YY} = g_{BB} = 0.72$ and  $M_{BB} = M_{B'B'} = 200$~GeV. 
 All off-diagonal parameters have been set to zero. Note that in
particular $M_{B B'}$ becomes sizable at the electroweak scale.

\section{Numerical results}
\label{sect:numerics}
All analytic expressions for masses, vertices, RGEs as well as 
one-loop corrections to the masses and tadpoles
were calculated using the \SARAH package
\cite{Staub:2008uz,Staub:2009bi,Staub:2010jh}.  For the
 generic expressions, those of ref.~\cite{Martin:1993zk} are used in the most
general form respecting the complete flavour structure. In addition,
gauge kinetic mixing effects in the RGEs are included using
the extensions of ref.~\cite{Fonseca:2011vn}. 
The loop corrections to all masses
as well as to the tadpoles are derived in \DRbar\ scheme and the 't Hooft
gauge. 

The numerical evaluation of the model is very similar to that of the default
 implementation of the
MSSM in \SPheno \cite{Porod:2003um,Porod:2011nf}: 
as the starting point, the SM gauge and Yukawa couplings
are determined using one-loop relations as given in
ref.~\cite{Pierce:1996zz} which are extended to our model.
 The vacuum expectation values \(v_d\) and \(v_u\) are
calculated with respect to the given value of \(\tan\beta\) at
\(M_Z\), while \(v_{\eta}\) and \(v_{\bar{\eta}}\) are derived from
the input values of $M_{Z'}$ and $\tan\beta'$ at the SUSY scale.

The RGEs for the gauge and Yukawa couplings are evaluated up to the
SUSY scale, where the input values of $Y_{\nu}$ and $Y_x$ are set.
Afterwards, a further evaluation of the RGEs up to the GUT scale takes
place. After setting the boundary conditions all parameters are
evaluated back to the SUSY scale. There, the one-loop-corrected SUSY
masses are calculated using on-shell external momenta. These steps are
iterated until the relative change of all masses between
two iterations is below $10^{-4}$.

\subsection{Benchmark Points}
For the numerical analysis in the following we have chosen five points
in the constrained parameter space 
which have distinct features. An overview of these points is given in
\TAB~\ref{tab:benchmark}. A typical feature is that $m_0$ has
to be in the TeV range to be consistent with the existing bounds
on the $Z'$-mass. 

\begin{table}[!h]
\centering
\begin{tabular}{|c|c c c c c|}
\hline
 & \boringBM & \lightHiggsBM & \higgsinoBM & \blinoBM & \bileptinoBM \\
\hline
 & \parbox[0pt][3em][c]{1.5cm}{\centering Bino LSP}
 & \parbox[0pt][3em][c]{1.5cm}{\centering Light Higgs}
 & \parbox{1.5cm}{\centering Higgsino LSP}
 & \parbox{1.5cm}{\centering \blino LSP}
 & \parbox{2cm}{\centering Bileptino LSP}\\
\hline
\multicolumn{6}{|c|}{Input} \\
\hline
$m_0$~[GeV]     & 1000 &   600 & 3500 & 3000 &  1000 \\
$M_{1/2}$~[GeV] & 1200 &  1400 & 1000 & 2500 &  1500 \\
$\tan\beta$     &   10 &    20 & 46   &   40 &    20 \\
$\sign(\mu)$    &    1 &     1 & 1    &    1 &     1 \\  
$A_0$~[GeV]     &-1000 & -1000 & 0    & 1500 & -1500 \\
$\sign(\mu')$   &    1 &     1 &  1   &   1  &     1 \\
$\tan\beta'$    & 1.07 & 1.055 & 1.34 & 1.20 &  1.15 \\
$M_{Z'}$~[GeV]  & 3000 &  2750 & 3600 & 2000 &  2500 \\
$Y_{x,11}$      & 0.41 &  0.43 & 0.42 & 0.42 &  0.37 \\
$Y_{x,22}$      & 0.41 &  0.43 & 0.42 & 0.42 &  0.40 \\
$Y_{x,33}$      & 0.41 &  0.43 & 0.36 & 0.42 &  0.40 \\
\hline
\multicolumn{6}{|c|}{CP-even Higgs sector}        \\
\hline
$m_{h_1}$~[GeV] & 110.1   &   14.7 &  122.9 &  124.1 &  123.9 \\
$m_{h_2}$~[GeV] & 124.2   &  123.6 &  849.2 &  273.9 &  208.1 \\
$m_{h_3}$~[GeV] & 1934.3  & 1877.3 & 2042.3 & 3008.7 &  2165.0 \\
$m_{h_4}$~[GeV] & 4044.1  & 3414.5 & 5914.8 & 6830.5 &  3007.8 \\
\hline
$|Z^H_{13}|^2+|Z^H_{14}|^2$ & 0.8471 & 0.9978 & 0.0002 & 0.0046 & 0.0026 \\
$|Z^H_{23}|^2+|Z^H_{24}|^2$ & 0.1529 & 0.0022 & 0.9998 & 0.9954 & 0.9973 \\
\hline
$\Gamma(h_{SM}) $~[MeV]                            & 2.22 & 2.47    & 2.43  & 4.15   & 2.47 \\
$\text{Br}(h_{SM} \to \gamma \gamma)\cdot 10^{-3}$ & 4.13 & 4.29    & 4.28  & 2.61   & 4.34 \\
\hline
\multicolumn{6}{|c|}{Neutralino sector} \\
\hline
$m_{\tilde{\chi}^0_1}$~[GeV] & 583.3  &  681.1  &  461.6  &    22.3 & 678.0 \\
$m_{\tilde{\chi}^0_2}$~[GeV] & 987.0  & 1150.9  &  501.8  &  1284.0 & 735.1 \\
$m_{\tilde{\chi}^0_3}$~[GeV] & 1501.0 & 1222.6  &  525.9  & 2025.9 & 1241.9 \\
$m_{\tilde{\chi}^0_4}$~[GeV] & 1508.0 & 1688.0  &  863.3  & 2063.0 & 1827.0\\
$m_{\tilde{\chi}^0_5}$~[GeV] & 1673.3 & 1694.0  &  1783.9 &  2148.0 & 1867.5  \\
$m_{\tilde{\chi}^0_6}$~[GeV] & 1967.0 & 1845.4  &  2672.5 & 3876.5 & 1871.5\\
$m_{\tilde{\chi}^0_7}$~[GeV] & 4139.3 & 3651.5  &  4876.2 & 4897.9 & 3131.4\\
\hline
$|Z^N_{11}|^2+|Z^N_{12}|^2$ & 0.9975 & 0.9975 & 0.4441 & 0.0137 & $O(10^{-5})$\\
$|Z^N_{13}|^2+|Z^N_{14}|^2$ & 0.0017 & 0.0013 & 0.5558 & $O(10^{-6})$ & $O(10^{-7})$\\
$|Z^N_{15}|^2$       & $O(10^{-5})$ & $O(10^{-5})$ & $ O(10^{-5})$ & 0.7770 & 0.0032 \\
$|Z^N_{16}|^2+|Z^N_{17}|^2$ & 0.0007 & 0.0012 & 0.0001 & 0.2092 & 0.9967\\
\hline
\end{tabular}
\caption{Points with distinct features: \boringBM is similar to the MSSM with a
 bino LSP, \lightHiggsBM provides a very light, bilepton-like Higgs boson,
 \higgsinoBM has a very large \higgsino fraction, \blinoBM has a \blino LSP and
 sizable mixing between the doublets and bileptons in the Higgs sector, and
 \bileptinoBM has a bileptino LSP. $h_{SM}$ denotes the Higgs fields which is 
most similar to the SM Higgs particle.}
\label{tab:benchmark}
\end{table}

\boringBM provides a neutralino LSP similar to the MSSM and also the lightest
 scalar Higgs is very similar to that of the MSSM and serves mainly
 to exemplify technical details. \lightHiggsBM demonstrates that it is
 possible to increase the Higgs mass of the MSSM-like light Higgs boson through
 a mixing with the new \BL fields. \higgsinoBM has an LSP with a large \higgsino
 fraction that is difficult to reach within the MSSM. \blinoBM and \bileptinoBM
 show two new dark matter candidates: a \blino and a bileptino LSP. 

We have also chosen points which lead to SM-like Higgs
masses in the preferred range of 123-126~GeV \cite{Chatrchyan:2012tx,:2012si}.
 As we will show
in the following, the extended Higgs sector has an impact also the the MSSM
 Higgs masses.
For instance, using the CMSSM parameters of \boringBM, the light Higgs would
 have a mass of
122.2~GeV, \IE~1.5~GeV lighter. However, this effect will be smaller for points
like \higgsinoBM or \blinoBM for which a larger mass splitting between the Higgs
 fields of
both sectors is present. In this case, the light Higgs mass in the \BLSSM agrees
 with that of the MSSM for the same parameters. One can also see that the
 branching ratios to 
two photons are always larger than in the SM except for \blinoBM, where
 the additional
neutralinos are so light that
 Br$(h \to \tilde{\chi}_1^0 \tilde{\chi}_1^0) = 43.6\%$. This feature
 significantly softens, of course, the bounds on the Higgs mass.

\subsection{Precision of the mass calculation: the impact of kinetic mixing}
\begin{table}[!H]
\centering
\begin{tabular}{|c|c|c|c||c|c|c|}
 \hline
particle & MSSM${}^{1L}$ & \BL${}^{1L}_{NKM}$ & \BL${}^{1L}_{KM}$
 & MSSM${}^{2L}$ & B-L${}^{2L}_{NKM}$ & B-L${}^{2L}_{KM}$ \\
\hline
$\tilde{d}_1$                   &        2280.8       &   2272.6     & 2273.2  &    2244.7     &  2234.6     &  2235.3  \\
$\tilde{d}_2$                   &        2429.8       &   2442.3     & 2449.9  &    2403.4     &  2414.7     &  2421.8  \\
$\tilde{d}_{3,4}$               &        2445.5       &   2457.9     & 2465.5  &    2418.7     &  2430.0     &  2437.0  \\
$\tilde{d}_{5,6}$               &        2570.9       &   2563.2     & 2564.2  &    2519.4     &  2509.7     &  2510.9  \\
$\tilde{u}_1$                   &        1832.5       &   1848.5     & 1834.0  &    1828.7     &  1842.4     &  1828.9  \\
$\tilde{u}_2$                   &        2301.3       &   2294.1     & 2294.2  &    2266.7     &  2257.8     &  2258.1  \\
$\tilde{u}_{3,4}$               &        2459.3       &   2471.4     & 2461.3  &    2428.9     &  2439.1     &  2429.6  \\
$\tilde{u}_{5,6}$               &        2569.7       &   2561.9     & 2563.0  &    2518.2     &  2508.5     &  2509.7  \\
$\tilde{e}_1$                   &        1087.2       &   1064.3     & 1056.3  &    1078.1     &  1043.4     &  1040.9  \\
$\tilde{e}_{2,3}$               &        1103.1       &   1080.4     & 1072.9  &    1093.8     &  1059.4     &  1057.3  \\
$\tilde{e}_4$                   &        1287.6       &   1379.7     & 1356.1  &    1259.9     &  1345.3     &  1323.6  \\
$\tilde{e}_{5,6}$               &        1293.4       &   1385.5     & 1362.1  &    1265.6     &  1351.0     &  1329.4  \\
$\tilde{\nu}^R_{1,2,3}$         &         -           &   848.6      & 698.8   &        -      &  944.3      &  809.1  \\
$\tilde{\nu}^R_4$               &         -           &   1376.7     & 1353.1  &       -       &  1342.2     &  1320.4  \\	
$\tilde{\nu}^R_{5,6}$           &         -           &   1382.9     & 1359.5  &       -       &  1348.3     &  1326.7  \\
$\tilde{\nu}^{(I)}_1$           &        1284.2       &   1376.7     & 1353.1  &      1256.3   &  1342.2     &  1320.4  \\
$\tilde{\nu}^{(I)}_{2,3}$       &        1290.6       &   1382.9     & 1359.5  &      1262.8   &  1348.3     &  1326.7  \\
$\tilde{\nu}^I_{4,5,6}$         &         -           &   3321.5     & 3220.8  &       -       &  3307.8     &  3205.0  \\
$h_1$                           &        121.9        &   121.6      & 123.3   &      122.1    &  121.8      &  110.1  \\
$h_2$                           &         -           &   127.2      & 102.2   &       -       &  130.9      &  124.2 \\
$h_3$                           &        1937.0       &   1934.1     & 1920.4  &       1952.2  &  1948.3     &  1934.3  \\
$h_4$                           &         -           &   4111.0     & 4109.3  &       -       &  4045.4     &  4044.1  \\
$A^0$                           &        1937.5       &   1936.1     & 1922.3  &      1952.6   &  1950.2     &  1936.2  \\
$A^0_\eta$                      &         -           &   2829.0     & 2820.3  &       -       &  2733.9     &  2725.5  \\
$H^+$                           &        1941.0       &   1938.2     & 1924.4  &      1956.1   &  1952.3     &  1938.3  \\
$\tilde{g}$                     &        2669.9       &   2670.2     & 2670.2  &      2602.1   &  2600.9     &  2600.7  \\
$\nu_{4,5,6}$                   &         -           &   1046.9     & 1046.5  &       -       &  987.6      &  987.2  \\
$\tilde{\chi}^+_1$              &        1046.3       &   1477.5     & 1467.4  &     988.5     &  1518.4     &  1508.2  \\
$\tilde{\chi}^+_2$              &        1480.7       &   3023.5     & 3020.4  &     1522.6    &  3023.2     &  3020.1  \\
$\tilde{\chi}^0_1$              &        550.3        &   550.3      & 598.4   &      534.8    &  533.9      &  583.3  \\
$\tilde{\chi}^0_2$              &        1046.2       &   1046.8     & 1046.3  &      988.3    &  987.4      &  987.0  \\
$\tilde{\chi}^0_3$              &        1472.2       &   1468.9     & 1458.4  &      1515.7   &  1511.6     &  1501.0  \\
$\tilde{\chi}^0_4$              &        1480.5       &   1477.3     & 1467.3  &      1522.3   &  1518.2     &  1508.0  \\
$\tilde{\chi}^0_5$              &         -           &   1749.0     & 1729.0  &       -       &  1695.2     &  1673.3  \\
$\tilde{\chi}^0_6$              &         -           &   1959.6     & 1947.5  &       -       &  1978.0     &  1966.9  \\
$\tilde{\chi}^0_7$              &         -           &   4175.5     & 4183.5  &       -       &  4131.2     &  4139.3  \\
\hline
\end{tabular}
\caption{Comparison of mass spectrum in the MSSM and the BLSSM for 
input parameters \boringBM of \TAB~{\protect \ref{tab:benchmark}}. 
We give the masses for one- and two-loop RGE-evaluations. 
In addition, we include a comparison of the case of properly taking
into account gauge kinetic mixing (KM) versus neglecting it
(NKM).
}
\label{tab:MSSMvsBL}
\end{table}

\begin{figure}[!h]
\begin{minipage}{\linewidth}
\includegraphics[width=0.49\linewidth]{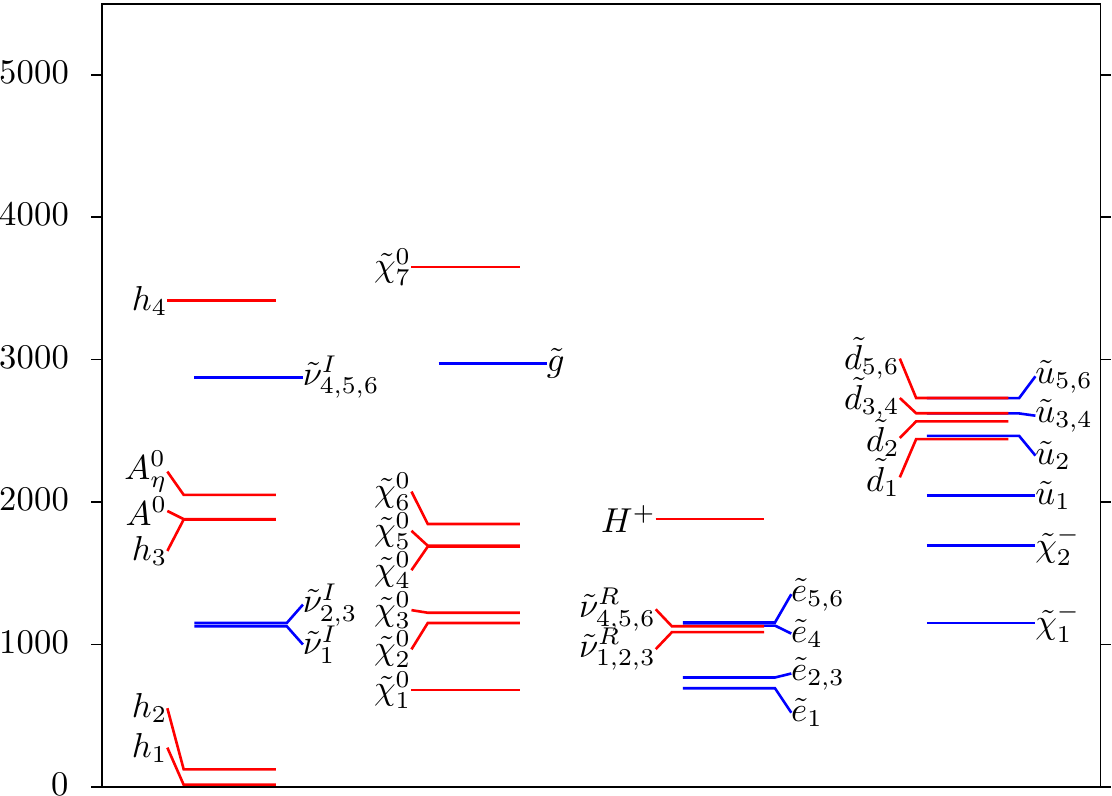} 
\hfill 
\includegraphics[width=0.49\linewidth]{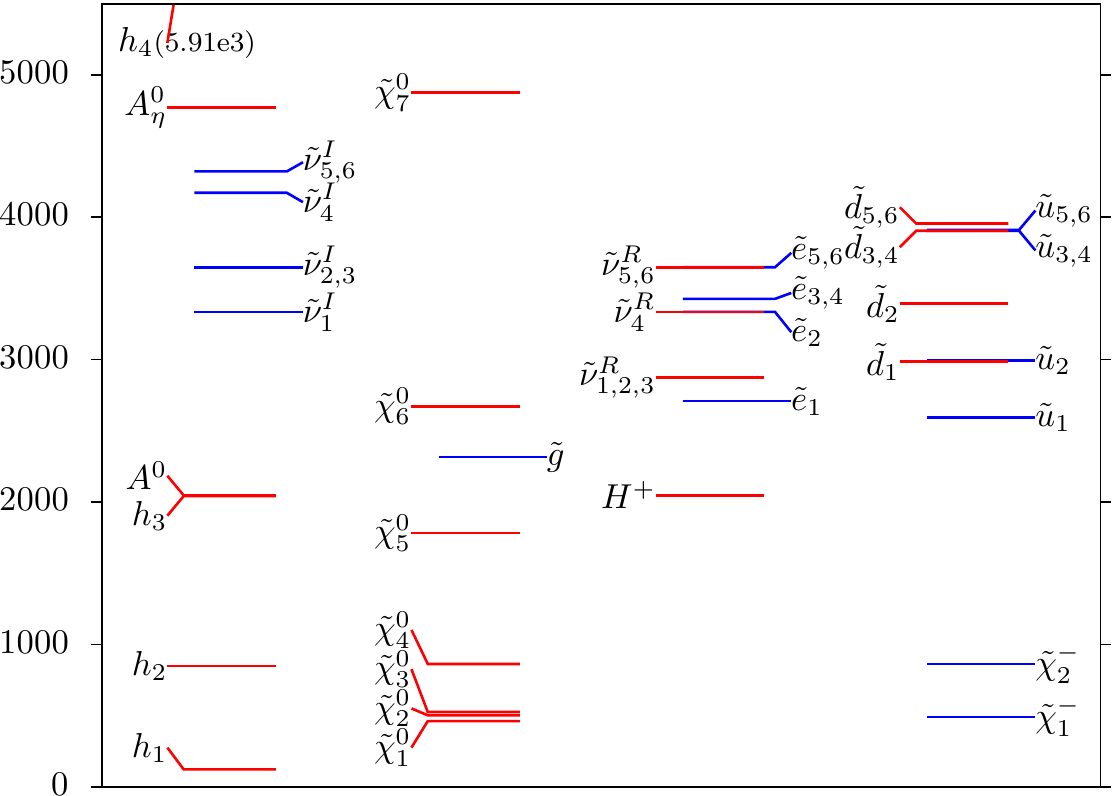}  \\
\includegraphics[width=0.49\linewidth]{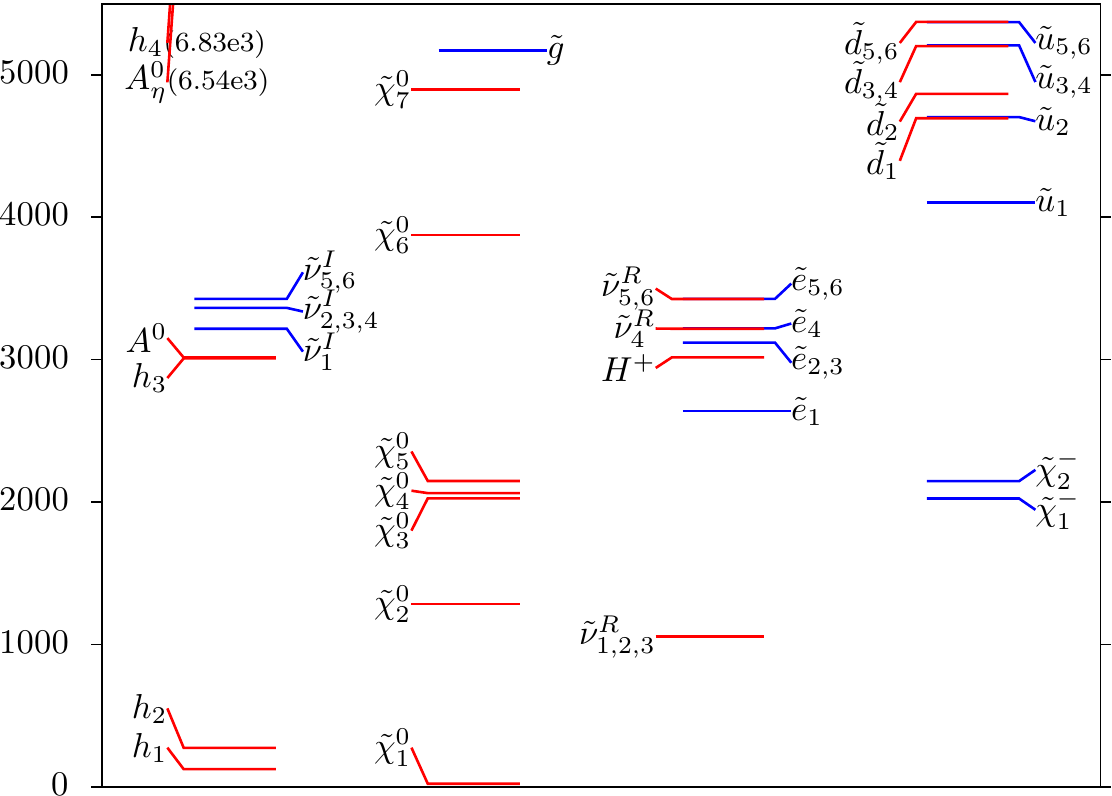} 
\hfill
\includegraphics[width=0.49\linewidth]{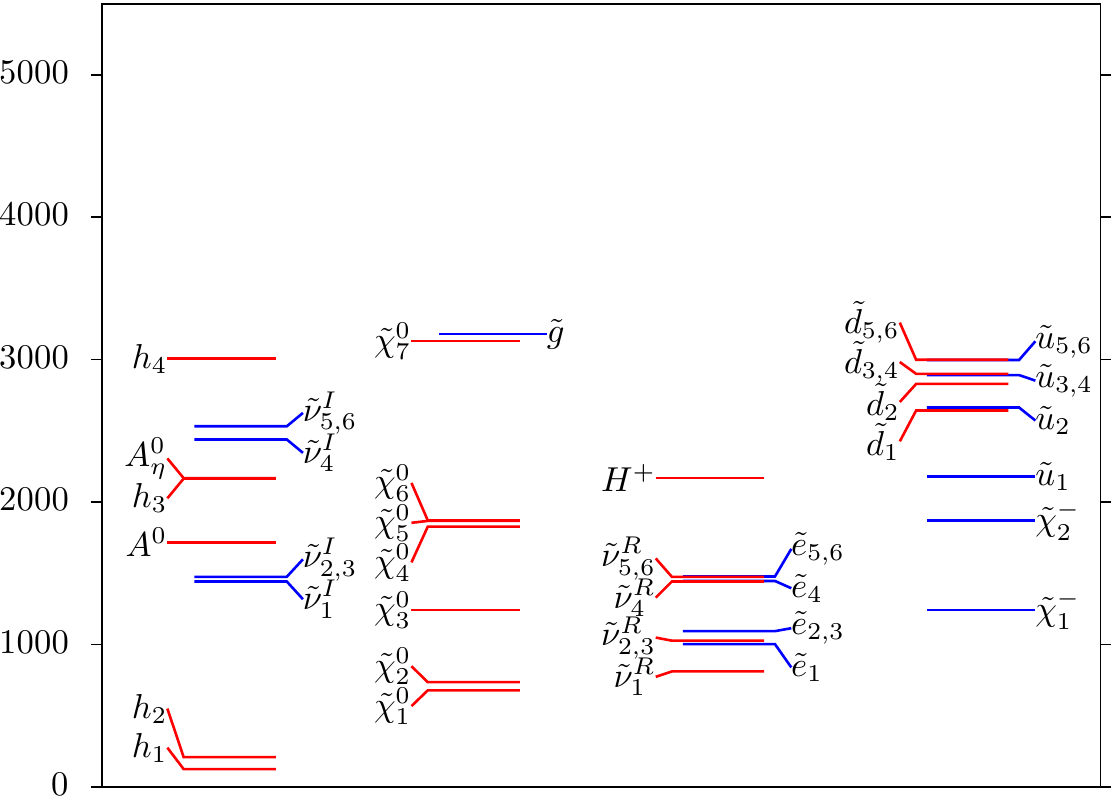} 
\caption{Spectrum of \lightHiggsBM (upper  left figure), \higgsinoBM (upper right figure), \blinoBM (lower left figure) and \bileptinoBM (lower right figure). All masses are given in GeV. The masses of the neutralinos and scalar Higgs fields are also given in table~\protect{\ref{tab:benchmark}}. }
\end{minipage}
\end{figure}

\begin{figure}[hbt]
 \begin{minipage}{\linewidth}
\includegraphics[width=0.48\linewidth]{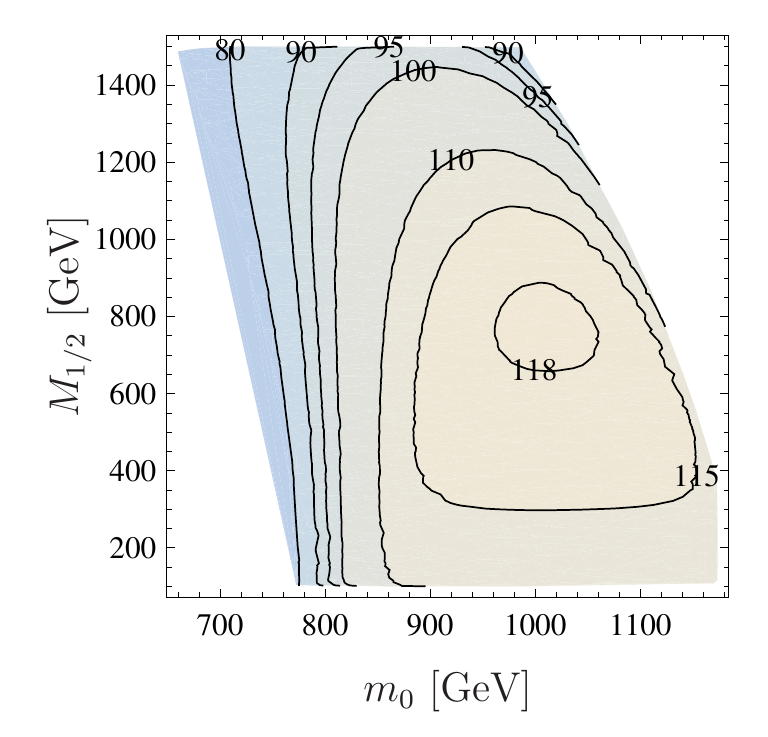} 
\hfill
\includegraphics[width=0.48\linewidth]{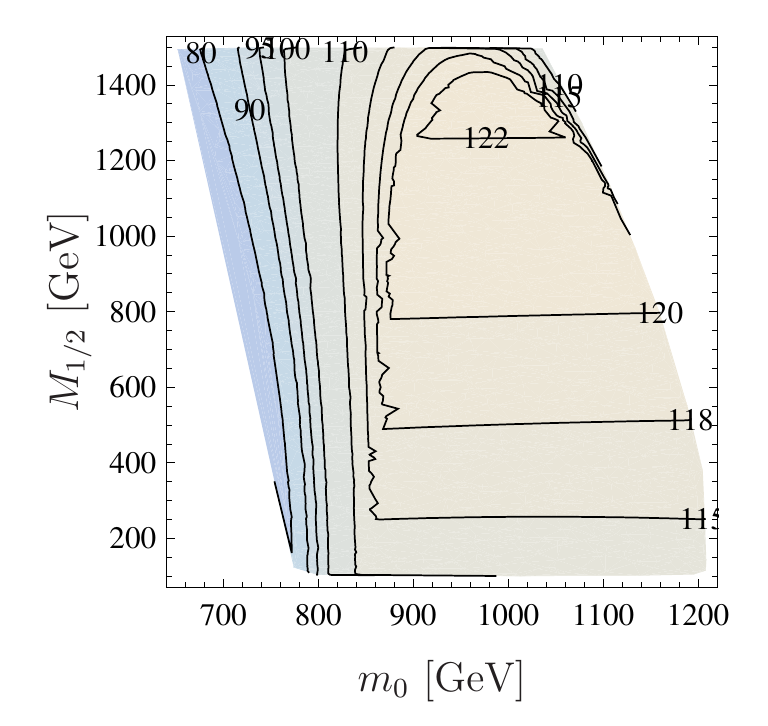} \\
 \includegraphics[width=0.48\linewidth]{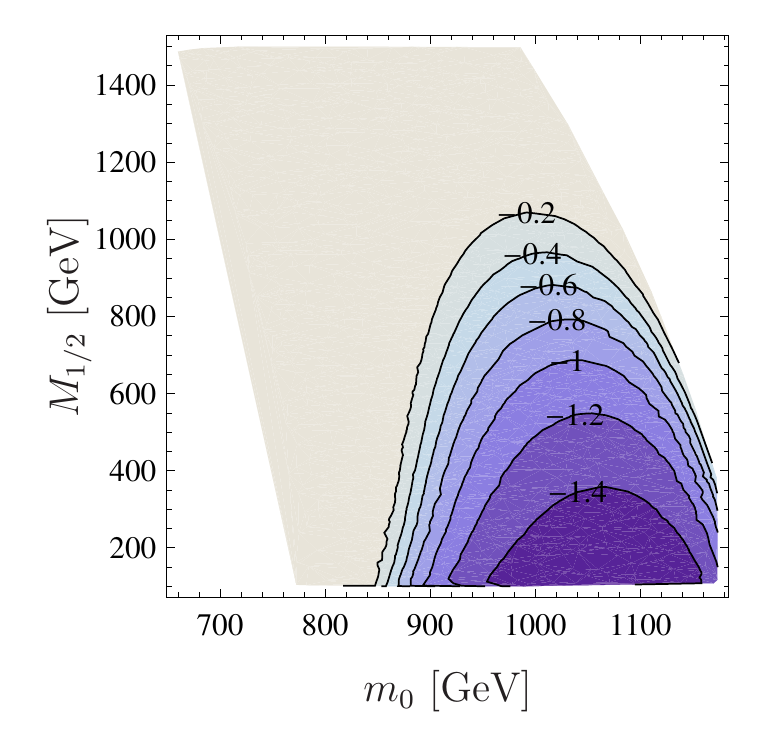} 
\hfill
 \includegraphics[width=0.48\linewidth]{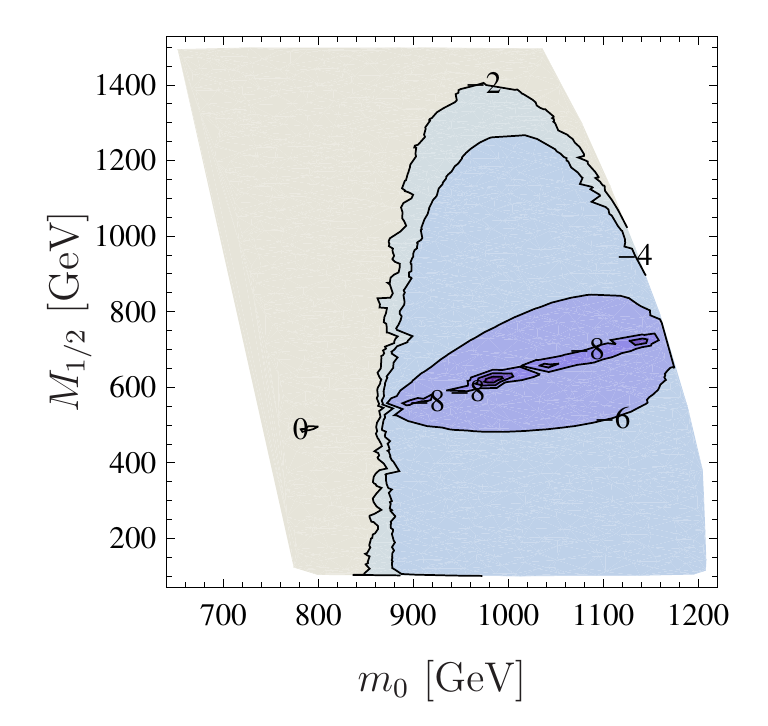} 
   \end{minipage}
\caption{Mass of the lightest Higgs (first row) and the logarithm of the \blino
 fraction of the lightest Higgs (second row). The other parameters are those of
 \boringBM. First left: with kinetic mixing, right: without kinetic mixing.}
\label{fig:Higgs_KM}
\end{figure}

Before we take a closer look on the Higgs and neutralino sector,
we want to comment on the precision of the mass calculation. As already
 mentioned, we use two-loop RGEs and one-loop corrections to the masses. In
 addition, we take the feature
of kinetic mixing into account which is often neglected in literature. 
To show the importance of the kinetic mixing and to compare the
resulting mass spectrum of the \BL
model with the MSSM, we show in \TAB~\ref{tab:MSSMvsBL} the masses 
calculated at the one-loop level using one-
or two-loop RGEs. For an easier comparison of the various effects, 
we fix the gauge the GUT scale to $2 \cdot 10^{16}$~GeV and the 
gauge couplings at the GUT scale to $g_1 = g_2 = g_3 = g_{BL} = 0.72$.

It can be seen that the most pronounced differences are
in the neutralino and Higgs sectors. In case of sfermions
and the heavier Higgs bosons
the differences between the MSSM and BLSSM particle
spectrum is about 1-2 percent and thus relatively small.
 They are larger in the charged slepton
sector since here the additional \UBL leads to larger effects. 
The main reason
for the smallness of the differences is the required largeness of 
$m_0$. Moreover,
the $D$-term effects due to the extra \UBL
are also small as $\tan\beta'$ is close to 1. 
In the sneutrino sector the scalar and pseudoscalar particles
behave quite differently: while the mass shifts in the pseudoscalar
sector are rather  moderate,
the masses of the scalars change by more than 100~GeV. 
The cause for this is a partial  cancellation for the given parameter 
point between the large positive terms in \EQ~(\ref{eq:mRR}), 
${m_\nu^2} +v_{\eta}  {Y_x  Y_x^*}$, and the large negative term 
$-4 \sqrt{2} Y_x {\mu'}^*$ which is very sensitive to the exact values of
all parameters at the SUSY scale. This also implies that
 kinetic mixing  is particularly important for this sector and
could even trigger  $R$-parity violation \cite{workinprep2}.  

In this table we also demonstrate that neglecting the
gauge kinetic effects can have dramatic effects in the
extended Higgs sector: the mass of $h_2$ state would 
be predicted to be about $20\%$ larger.

Also in the neutralino sector there is quite some impact
for the LSP masses which would be wrong by about $10\%$
if gauge kinetic terms were to be neglected. But also the 
properties of the lightest Higgs particle can change: we show 
in \FIG~\ref{fig:Higgs_KM} a comparison between the mass and bilepton 
fraction of the lightest with and without kinetic mixing. 
It can be seen that the masses are clearly shifted while, of
course, there is also huge difference of several orders in the bilepton
fraction between both cases. While the bilepton contribution for
MSSM-like scalars in the case without kinetic mixing is solely based
on the mixing at one-loop level, the off-diagonal gauge couplings
introduce already a tree-level mixing.

\begin{figure}[t]
 \begin{minipage}{0.99\linewidth}
\centering
\includegraphics[width=0.48\linewidth]{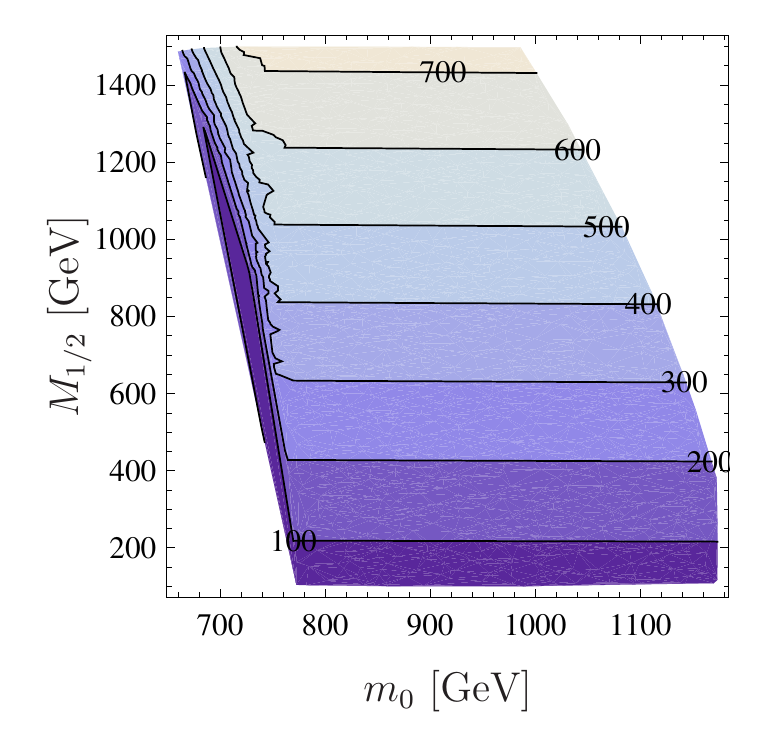}
\hfill
\includegraphics[width=0.48\linewidth]{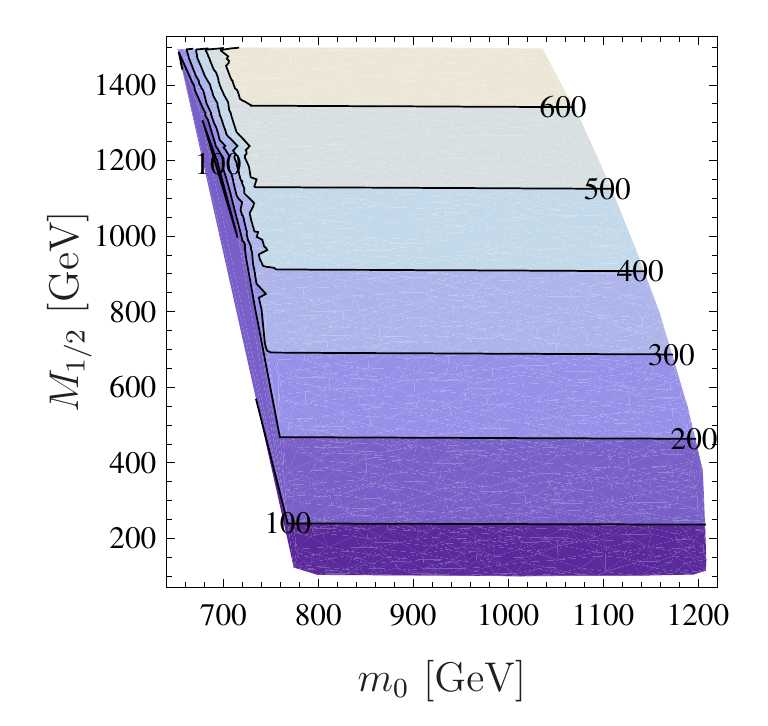}  \\
 \includegraphics[width=0.48\linewidth]{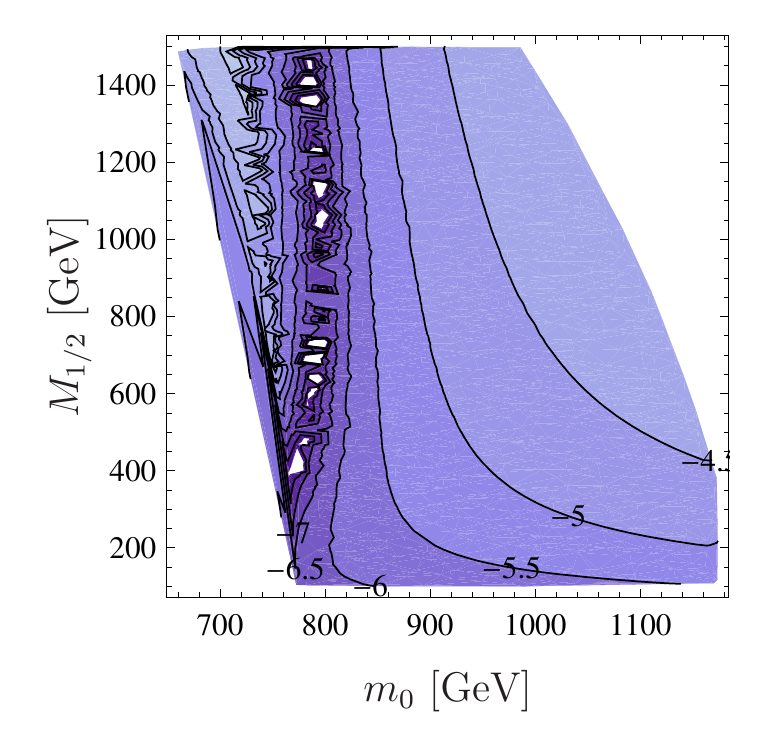} 
\hfill
\includegraphics[width=0.48\linewidth]{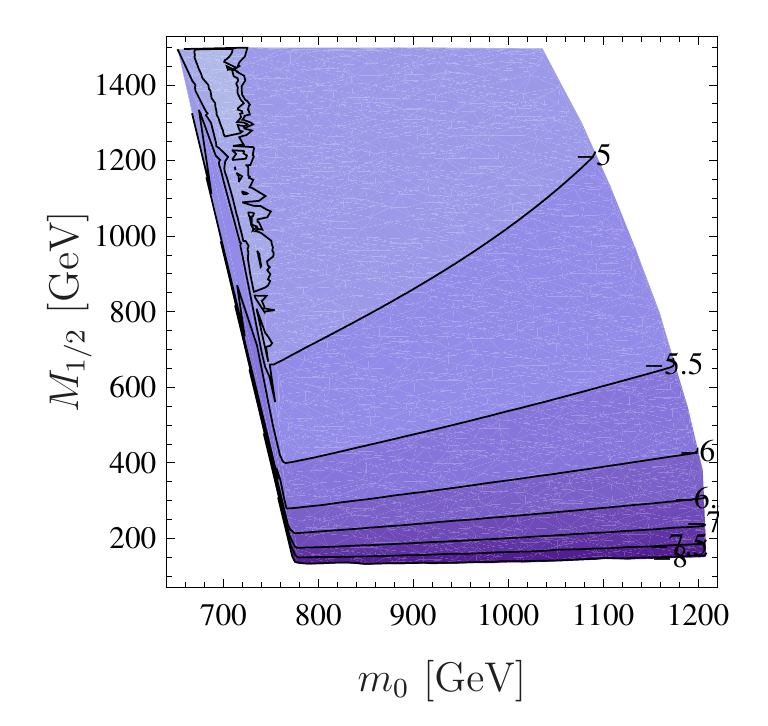} 
   \end{minipage}
\caption{The $(m_0,M_{1/2})$ plane for a bino-LSP with kinetic mixing (left) and
 without (right). First row: mass of lightest neutralino. Second row: \blino
 fraction. The other parameters correspond to point \boringBM.}
\label{fig:Neutralino_KM}
\end{figure}

We turn now to to the neutralino sector. 
When the lightest neutralino is bino- or \higgsino-like in this model,
it shares the common features of the analogous neutralino in the MSSM.
 However,
the masses differ at the SUSY scale for the same GUT scale parameters 
mainly as a consequence of gauge kinetic mixing in the gaugino sector
which in our example amounts to shift of about 10 per-cent.
This effect is
especially important in the case of a bino LSP because the off-diagonal
gaugino mass parameter $M_{B B'}$ can easily reach values of
$\mathscr{O}(-50~\GeV)$ (for $M_{1/2} \simeq 200~\GeV$) or even
$\mathscr{O}(-150~\GeV)$ (for $M_{1/2} \simeq
1000~\GeV$).
For this
purpose, we depict in the left column of
\FIG~\ref{fig:Neutralino_KM} the mass of the lightest neutralino in
the $(m_0,M_{1/2})$-plane as well the \blino fraction. To point 
out again the importance of kinetic mixing, we
show the same plots in the right column of \ref{fig:Neutralino_KM}
without the kinetic mixing. A shift of masses is clearly visible and
also the admixture differs by several orders. The other parameters
are the same as for \FIG~\ref{fig:Higgs_KM}.

For completeness we note that the differences
between tree-level masses and loop-corrected masses
are of similar size for the MSSM-particles. The R-sneutrinos
receive somewhat larger loop corrections 
of about 3-4 per-cent compared to a few per-mille for the
L-sneutrinos. However, interesting effects
in the Higgs and neutralino sector can happen which we discuss in the next
subsections.

\subsection{The Higgs sector} 
\begin{figure}[hbt]
 \begin{minipage}{0.99\linewidth}
\includegraphics[width=0.48\linewidth]{figs_paper/m0_m12_mH1} 
\hfill
\includegraphics[width=0.48\linewidth]{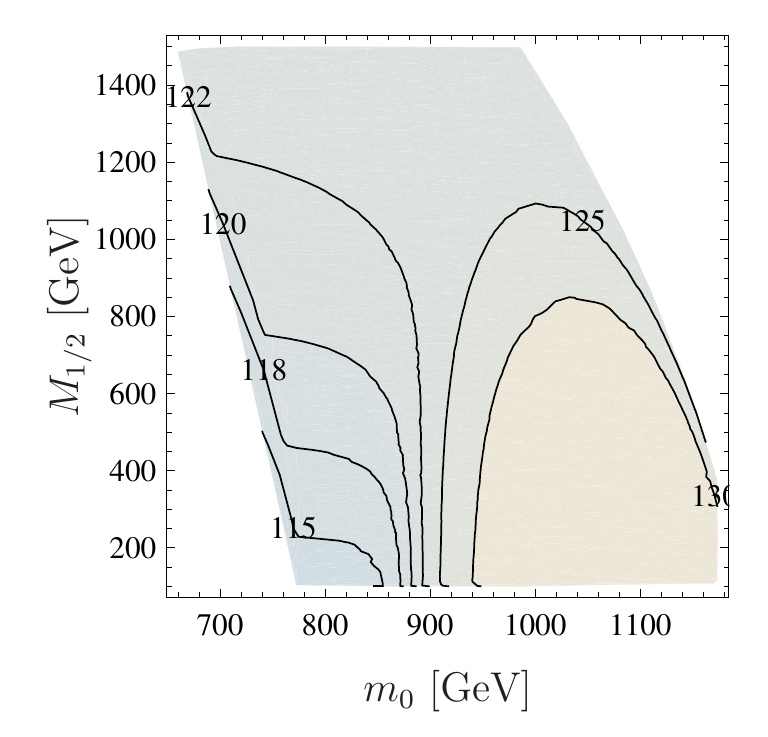} \\
 \includegraphics[width=0.48\linewidth]{figs_paper/m0_m12_H1-bilepton} 
 \hfill
\includegraphics[width=0.48\linewidth]{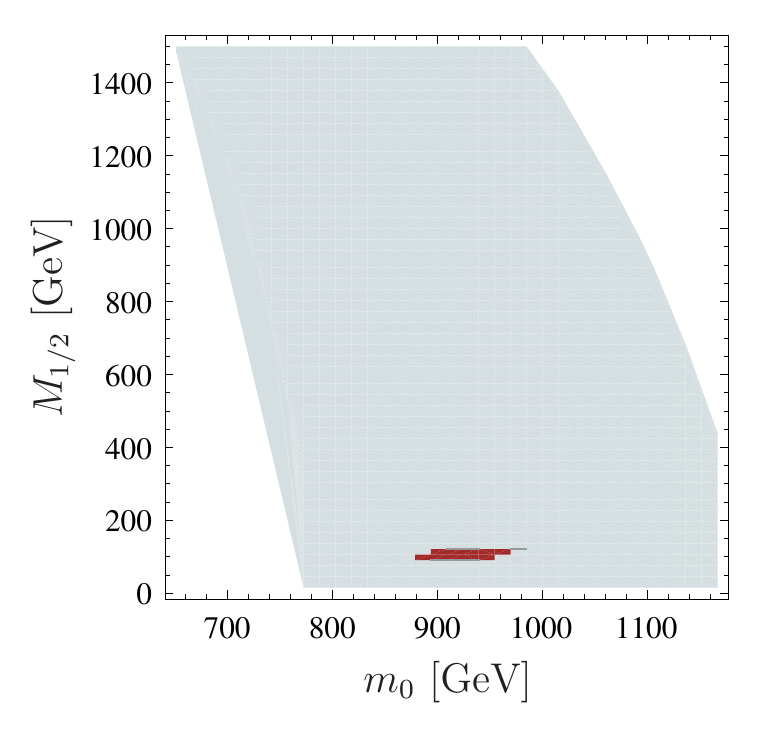}
   \end{minipage}
\caption{Mass of the lightest Higgs (first row on the left) and the second
 lightest Higgs (first row on the right), the logarithm of the bilepton fraction
 of the lightest Higgs (second row on the left) as well as the bounds from Higgs
 searches (second row on the right) in the $(m_0,M_{1/2})$-plane. For the Higgs
 search the saturation of the tightest bound as calculated by {\tt HiggsBounds} is 
 shown. While the blue is still consistent with all data, the red area is
 excluded. The most stringent channels are  $e^{+} e^{-} \to Z h_{1,2}, h_{1,2} \to b \bar{b}$  and 
 $p p \to h_{1,2} \to W^+ W^-$. The other parameters are as for point \boringBM. }
\label{fig:Higgs_Standard}
\end{figure}
In this section we  concentrate on the Higgs
sector and discuss new phenomenological aspects arising in the \BL
model. In \FIG~\ref{fig:Higgs_Standard} we show the lightest scalar
Higgs mass in the $(m_0,M_{1/2})$-plane. In addition, we give the bilepton
 fraction. The other
input values are the same as for \boringBM.  The nature of
the second lightest Higgs is roughly complementary to the lightest
one with respect to the ratio of the bilepton nature versus the
Higgs doublet nature. The reason is that the mixing between the
light states with the two heavy states, which have masses above
a TeV, is quite small. Note that complete region shown is compatible
with recent LHC data even though the mass of the second lightest Higgs boson is
in most parts above 140 GeV, as it is mainly a bilepton with
a small production cross section at the LHC.  
\begin{figure}[hbt]
 \begin{minipage}{0.99\linewidth}
\includegraphics[width=0.48\linewidth]{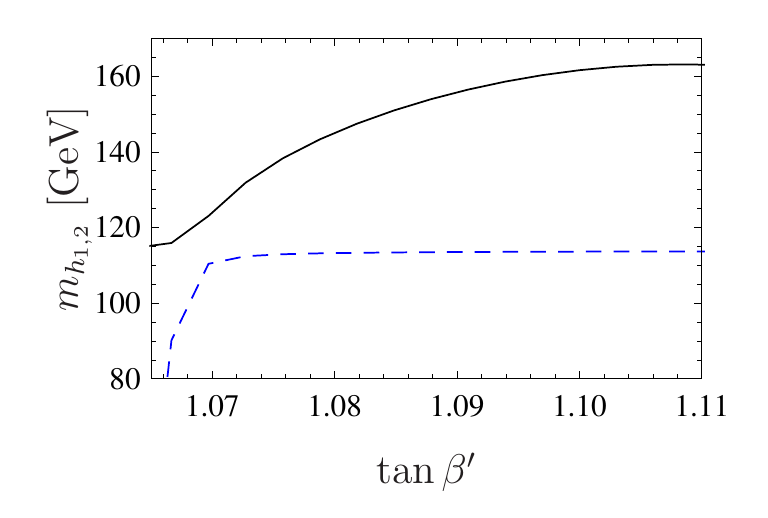} 
\hfill
 \includegraphics[width=0.48\linewidth]{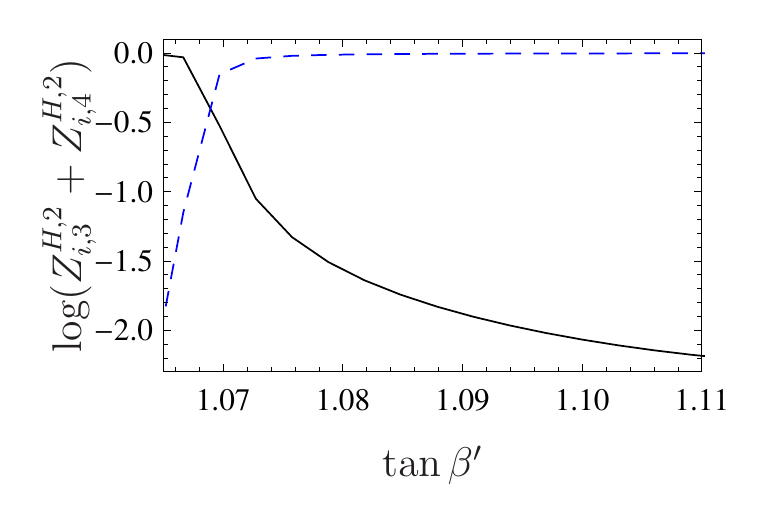} 
   \end{minipage}
\caption{a) masses of two lightest scalars. b) doublet (dashed blue) and bilepton
 (black) fraction of lightest higgs as function of $\tan\beta'$. The other input
 parameters are as for point \lightHiggsBM, but with $m_0 = 2 M_{1/2} = 1$~TeV }
\label{fig:Higgs_tbp}
\end{figure}

Close to the border of the allowed regions in the
 ($m_0,M_{1/2}$)-plane, the lightest 
Higgs particles become bilepton-like.
Not only can this be observed for a variation of $m_0$ and
$M_{1/2}$ but also by adjusting $\tan\beta'$, as shown in
\FIG~\ref{fig:Higgs_tbp} where we have fixed $m_0 = 1000~\GeV$ and
$M_{1/2} = 500~\GeV$. To understand this behavior we neglect
gauge kinetic mixing for simplicity as then the bilepton sector
decouples from the MSSM Higgs bosons. In this limit we obtain
at tree level a similar formula for  masses as for the MSSM Higgs bosons:
\begin{eqnarray}
m^2_{1,2} &=& \frac{1}{2} \left(
m^2_{Z'}+ m^2_{A^0_\eta} \mp 
\sqrt{(m^2_{Z'}+ m^2_{A^0_\eta})^2
     - 4 m^2_{Z'} m^2_{A^0_\eta} \cos^2(2 \beta')}\right)
\end{eqnarray}
Equations.~(\ref{eq:tadBmuP}) and (\ref{eq:mA2}) imply that
for fixed $M_{Z'}$,
 $m^2_{A^0_\eta}$ shows a sizable dependence on $\tan\beta'$.
 We checked that very light
bilepton-like Higgs scalars are not ruled out by experimental data
using {\tt HiggsBounds 3.6.1beta}
\cite{Bechtle:2008jh,Bechtle:2011sb}. However, the mixing between
the bilepton and the MSSM-like Higgs is rather small and thus
the branching ratio $h_2 \rightarrow h_1 h_1$ is at most a few
per-cent. Therefore, the main decay channels of the doublet
Higgs are still SM final states and the well-known bounds do hold.
However, as can be seen also in \FIG~\ref{fig:Higgs_tbp}, the mass of
the MSSM-like Higgs bosons gets pushed to larger values for very light
bilepton scalars. Such a behavior has already been observed
in the literature when considering models with extended gauge
symmetries 
\cite{Haber:1986gz,Drees:1987tp,Cvetic:1997ky,Zhang:2008jm,Ma:2011ea,
Hirsch:2011hg}.

\begin{figure}[t]
 \begin{minipage}{0.99\linewidth}
\includegraphics[width=0.48\linewidth]{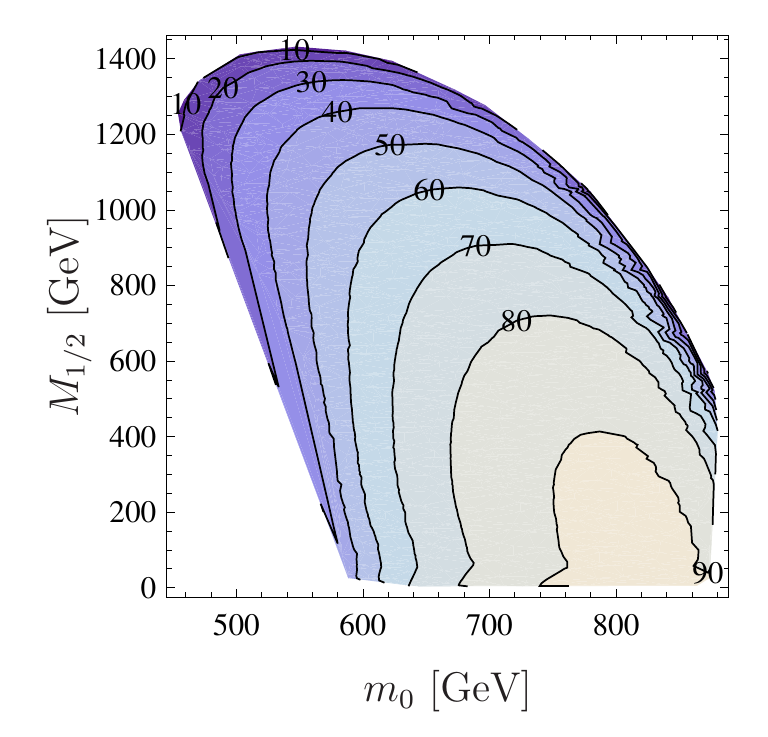}
\hfill
\includegraphics[width=0.48\linewidth]{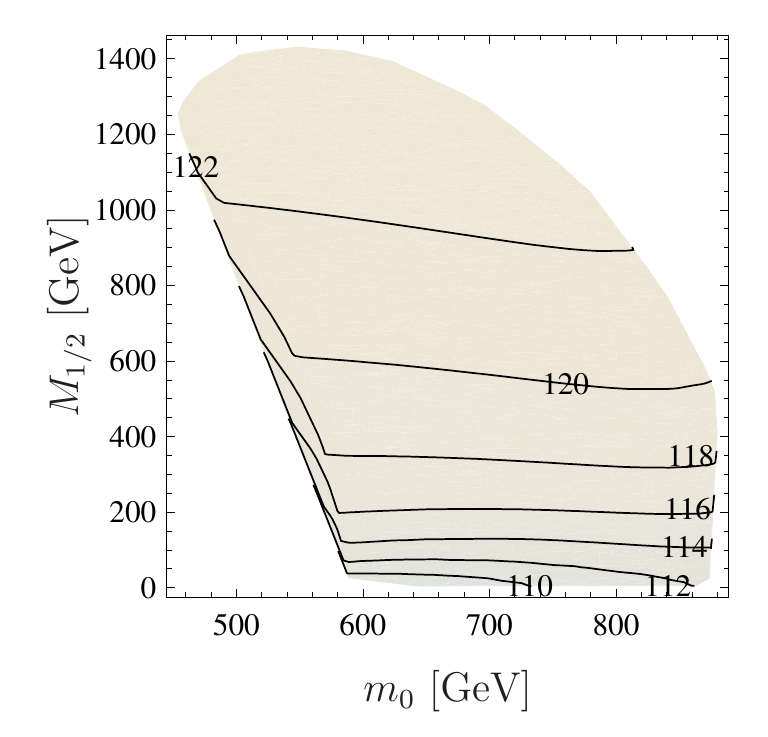} \\
\includegraphics[width=0.48\linewidth]{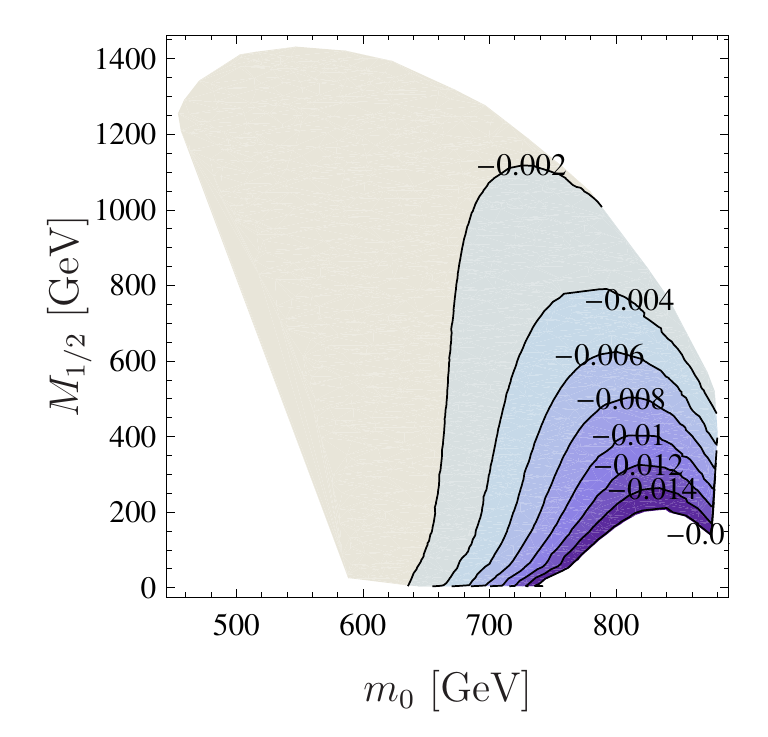}
\hfill
 \includegraphics[width=0.48\linewidth]{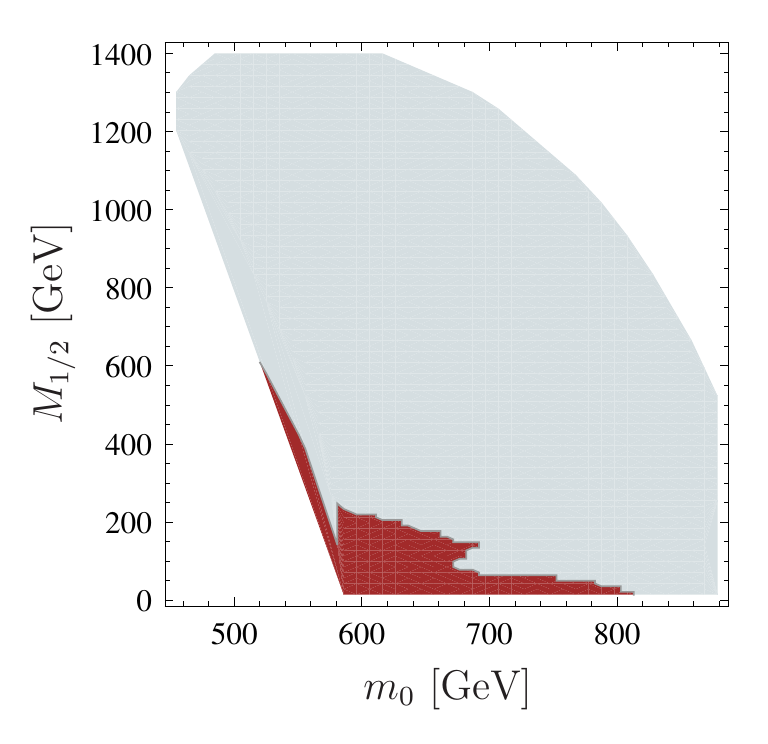}
   \end{minipage}
\caption{Mass of the two lightest Higgs fields (first row) as well as the
 logarithm of the bilepton fraction (left plot in second row) in the
 $(m_0,M_{1/2})$-plane. The right plot in the second row shows the saturation of
 the tightest bound as calculated by
 {\tt HiggsBounds}: the blue are is allowed, the red one excluded by Higgs
 searches. The most sensitive channels are $e^{+} e^{-} \to Z h_2, h_2 \to b \bar{b}$,
 $p p \to A^0 \to \tau \bar{\tau}$ and  $p p \to h_2 \to W^+ W^-$. 
 The other parameters are based on \lightHiggsBM.}
\label{fig:HeftyHiggs}
\end{figure}

In \FIG~\ref{fig:HeftyHiggs} we take the point \lightHiggsBM and vary
$m_0$ and $M_{1/2}$. We see that there is a sizable region
where the lightest Higgs, being essentially a bilepton, has a mass
of less than half of the second lightest, which is mainly
like the MSSM $h^0$. Even though the bilepton has only a small
admixture of the doublet Higgs bosons, it is large enough to
determine its main decay properties, which are mainly SM-like with
respect to its decay into SM fermions.

\paragraph*{Loop corrections}
Concerning the loop-corrections in the Higgs sector, the picture is often
 comparable with the MSSM: the
lightest two Higgs bosons receive very large corrections. These are even
larger for the chosen BLSSM point than in the MSSM. For the heavy
doublet scalar as well as the charged and MSSM-like pseudoscalar Higgs, 
the differences between tree-level and one-loop masses are rather
small and of the same size as in the MSSM. The neutrino Yukawa
couplings don't play any role in this context because the correct
explanation of neutrino data requires them to be
very small. However, there are sizable loop corrections due to the 
large $Y_x$ couplings similar to the top-stop contributions to the
 lighter MSSM-Higgs.
For example, in the case of the \blinoBM scenario, $m_{h_2}$ gets shifted 
from about
252 GeV at tree level to about 210 GeV at the one-loop level.
This is a consequence of the mass hierarchy
between fermions and bosons in the extended gauge/Higgs sector.

\begin{figure}[t]
 \begin{minipage}{0.99\linewidth}
\includegraphics[width=0.48\linewidth]{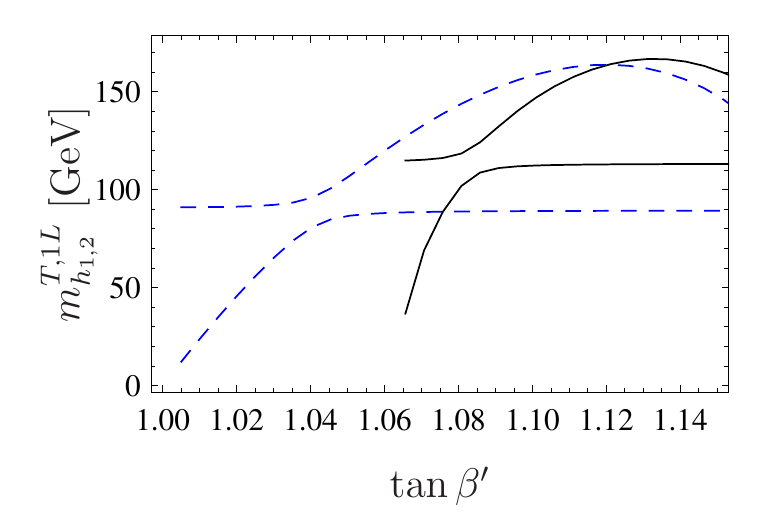} 
\hfill
 \includegraphics[width=0.48\linewidth]{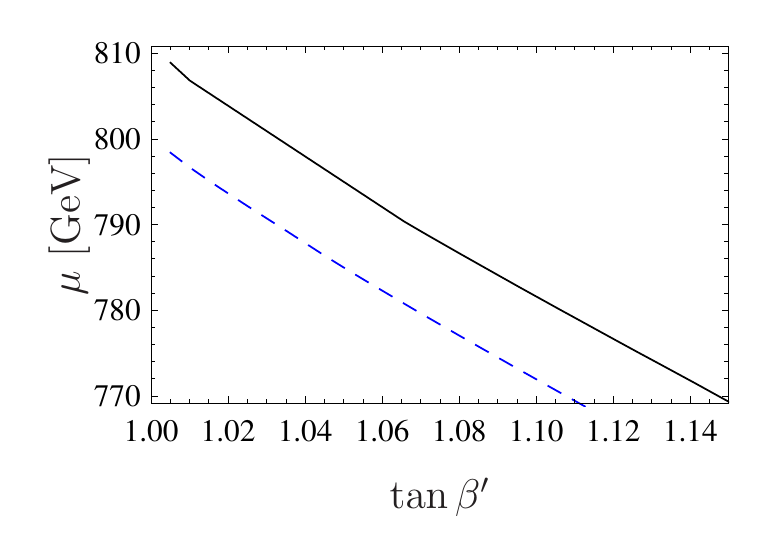} 
   \end{minipage}
\caption{Left: mass of the lightest Higgs fields at one-loop level (black) and
 tree level (dashed blue). Right: value of $\mu$ calculated from the one-loop corrected
 tadpole equations (black) and the tadpole equations at tree level (dashed blue). 
 The other input parameters correspond to those of \FIG~\protect{\ref{fig:Higgs_tbp}}}
\label{fig:tbp_H1H2_loop-tree}
\end{figure}

For completeness we note that in the Higgs sector one finds
that the mass of the bilepton-like Higgs field vanishes at tree level
in the limit
$\tan\beta' \to 1$. Note, however, that $\tan\beta'= 1$ is a saddle point
and not a minimum of the tree-level potential. This in turn implies
that the loop corrections will be large compared to the tree level
similar to as it is in the MSSM when considering there the limit
 $\tan\beta \to 1$.
We explicitly demonstrate this in \FIG~\ref{fig:tbp_H1H2_loop-tree}
where we compare the tree-level and one-loop masses of the two lightest
Higgs fields, fixing the input parameters as in point \lightHiggsBM 
but varying $\tan\beta'$. This behavior is also reflected in the 
values of $\mu$ calculated from the one-loop and tree-level tadpole equations
 as shown in the right plot of 
\FIG~\ref{fig:tbp_H1H2_loop-tree}. It can
 be understood from \EQ~(\ref{eq:tadmu}): compared to $\cos(2\beta')$,
all parameters and the one-loop corrections show only a very mild
dependence on  $\beta'$ as $\tan\beta'$ has to be close to one.
Therefore, the one-loop correction to the tadpole equation can
 be included in the first term and  effectively be 
absorbed in a redefintion of $\beta'$, denoted by  $\tilde{\beta}'$,
using the equation
 $\tilde{g} g_{BL} x^{2} \cos(2 {\beta'})
 + \frac{4 (\delta t_u - \delta t_d)}{\sec(2\beta)}
 = \tilde{g} g_{BL} x^{2} \cos(2 {\tilde{\beta}'})$ with a shifted
 $\tilde{\beta}'$.

\begin{figure}[th]
 \begin{minipage}{0.99\linewidth}
\includegraphics[width=0.48\linewidth]{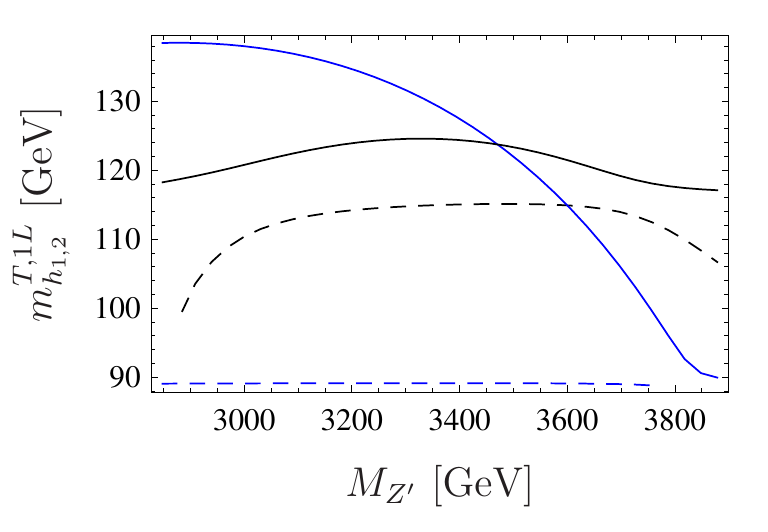} 
\hfill
 \includegraphics[width=0.48\linewidth]{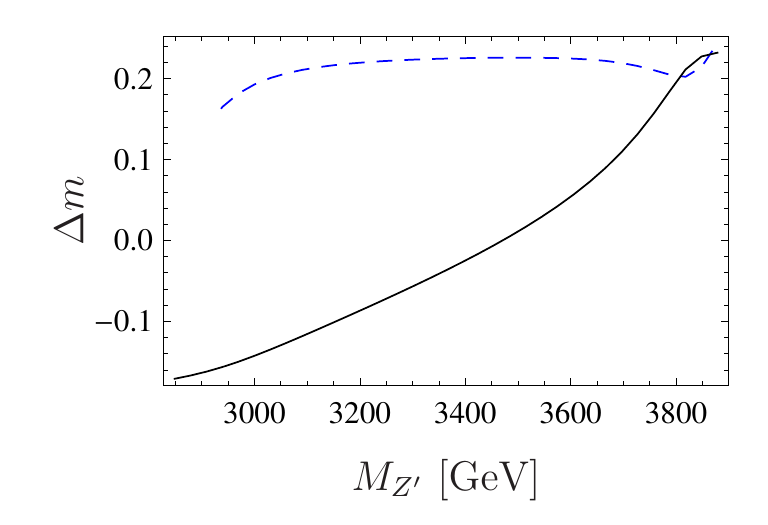} 
   \end{minipage}
\caption{Left: masses of the lightest Higgs (dashed) and second lightest Higgs
 (solid) at one-loop level (black) and tree level (blue) for a variation of
 $M_{Z'}$. Right: relative difference between tree-level and one-loop mass for
 the lightest Higgs (dashed blue)  and second lightest Higgs (solid black). The other input
 parameters correspond to point \boringBM.}
\label{fig:MZp_H1H2_loop-tree}
\end{figure}

In \FIG~\ref{fig:MZp_H1H2_loop-tree} we show the masses of the two lightest
 Higgs fields for \boringBM with a variation of $M_{Z'}$. As expected the
 tree-level
 and
 one-loop mass of the lightest Higgs which is consists mainly of the $SU(2)$
 doublet is nearly independent of $M_{Z'}$. In contrast to that, the tree-level
 mass of the bilepton-like Higgs depends strongly on $M_{Z'}$. Furthermore, the
 one-loop corrections can be nearly of the order known from the MSSM for the
 doublet Higgs depending on the mass of the $Z'$. Also the sign of the
 correction
 can change depending on the mass ordering of $Z'$ and the \blino-like
 neutralino.

\subsection{The neutralino sector}
\label{sec:neutralinos}
Similarly to how it is in the CMSSM, the lightest neutralino is often bino-like
and the main difference is, in this case, that the relation between
the parameters at different scales gets changed due to the gauge
kinetic mixing. Note that this
holds even though the soft-breaking gaugino mass term $M_{B'}$ is always
smaller than $M_1$, because, at one-loop level and without kinetic
mixing, the relation
\begin{equation}
 \frac{M_{1/2}}{g_{GUT}^2} = \frac{M_1}{g_Y^2} = \frac{M_{B'}}{g_{BL}^2}
\end{equation}
would hold and $g_{BL}$ is always smaller than $g_Y$ if unification
at the GUT scale is assumed, as can be seen in
\EQ~(\ref{eq:gammaMatrix}). However, usually there is a large
mixing between  the \blino with the bileptinos, leading
to heavy states. However, we will demonstrate that 
nevertheless regions
exist where the lightest neutralino is \blino- or even bileptino-like.
Therefore, a neutralino LSP can have four different natures in the \BLSSM
 (bino, \higgsino, \blino, bileptino) in contrast to only two
possibilities in the CMSSM. This can provide interesting features in the
context of dark matter \cite{workinprep}. 

\begin{figure}[t]
\begin{minipage}{0.99\linewidth}
\includegraphics[width=0.48\linewidth]{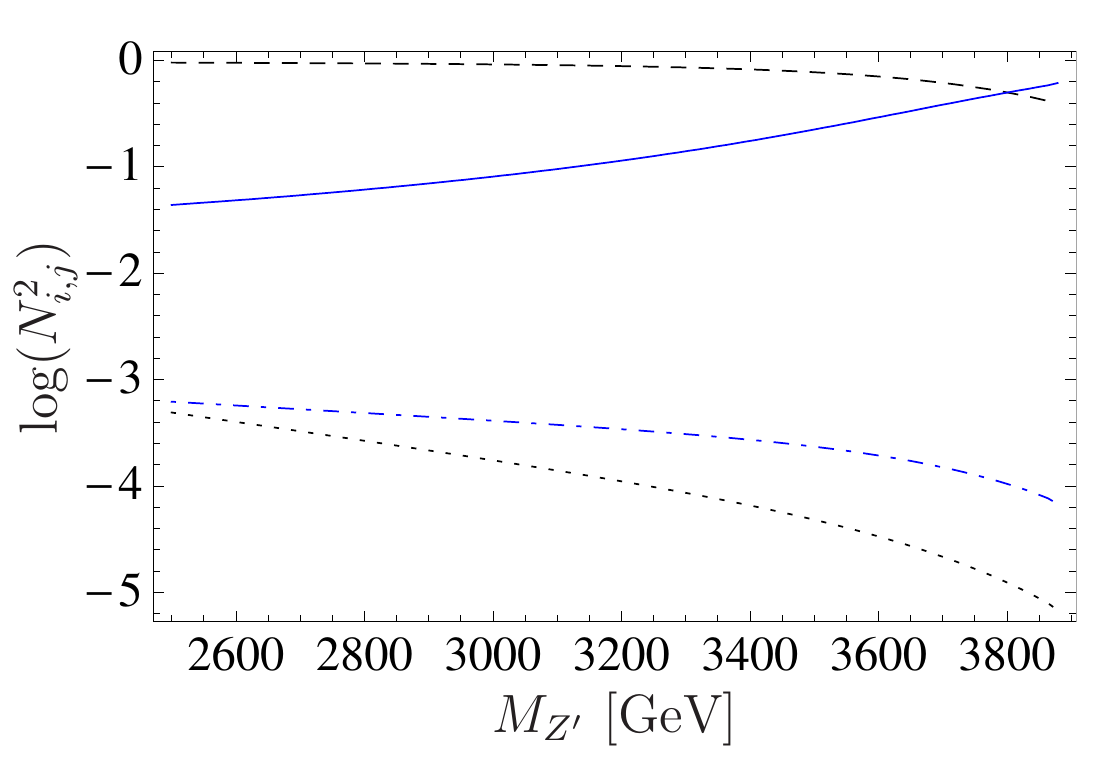} 
\hfill
\includegraphics[width=0.48\linewidth]{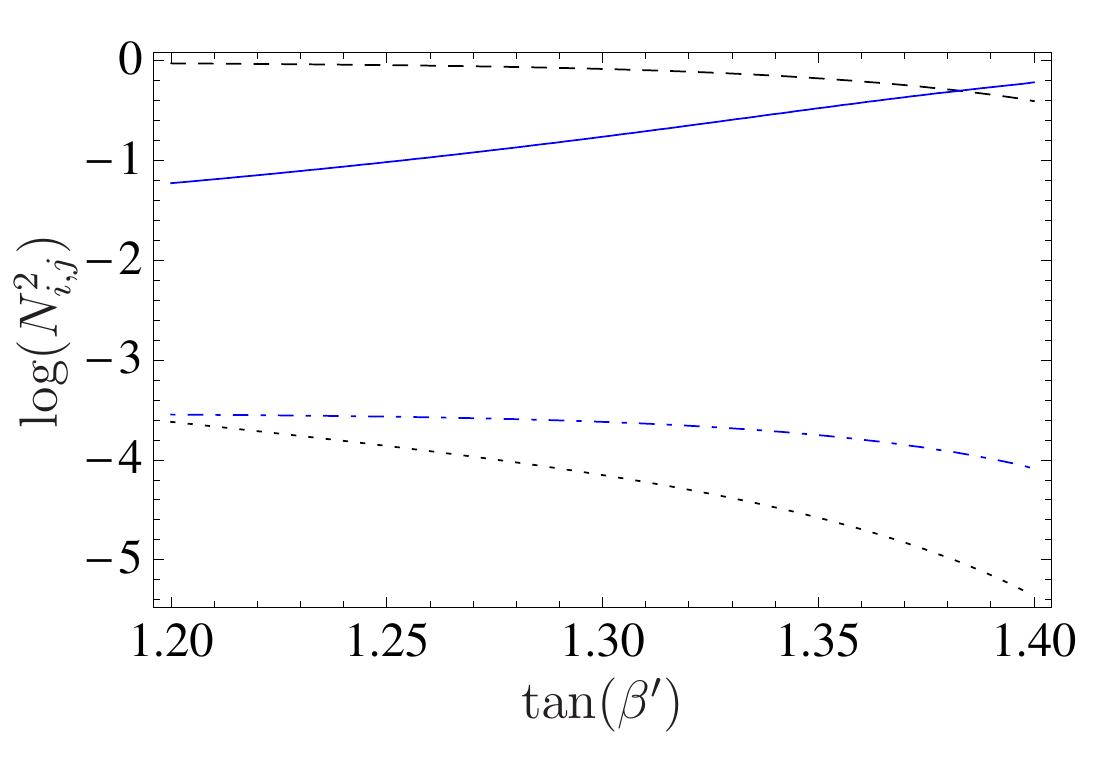}
\end{minipage}
\caption{The content of the LSP depending on the $\tan\beta'$ (right) and
 $M_{Z'}$ (left). The other parameters are those of \higgsinoBM. The color code
 is
 as follows: gaugino fraction (dashed black), \higgsino fraction (blue), logarithm of
 the \blino fraction  (dotted black) and bileptino fraction (dot-dashed blue).}
\label{fig:Higgsino}
\end{figure}

It is well known that in the MSSM, it is very hard to reach \higgsino fractions
 of the LSP larger than 50 per-cent: 
 even in the focus point region this fraction
hardly ever exceeds 30 per-cent. Only in a tiny region where $|\mu|$ gets
close to 0, and which is excluded by LEP data, does it get larger than
50 per-cent.  In contrast in our model new contributions show
up in the formula for  $\mu$ in \EQ~(\ref{eq:tadmu}), in
particular the term $ \tilde{g} g_{BL} x^{2} \cos(2 {\beta'})$.
Using universal boundary conditions it  is in
 general negative and is particularly sizable for large
$\tan\beta'$ and $M_{Z'}$. 
Therefore it is possible to increase the \higgsino fraction of the
 LSP by increasing these two parameters with fixed CMSSM parameters as shown in
 \FIG~\ref{fig:Higgsino}.

\begin{figure}[t]
 \begin{minipage}{0.99\linewidth}
\centering
 \includegraphics[width=0.48\linewidth]{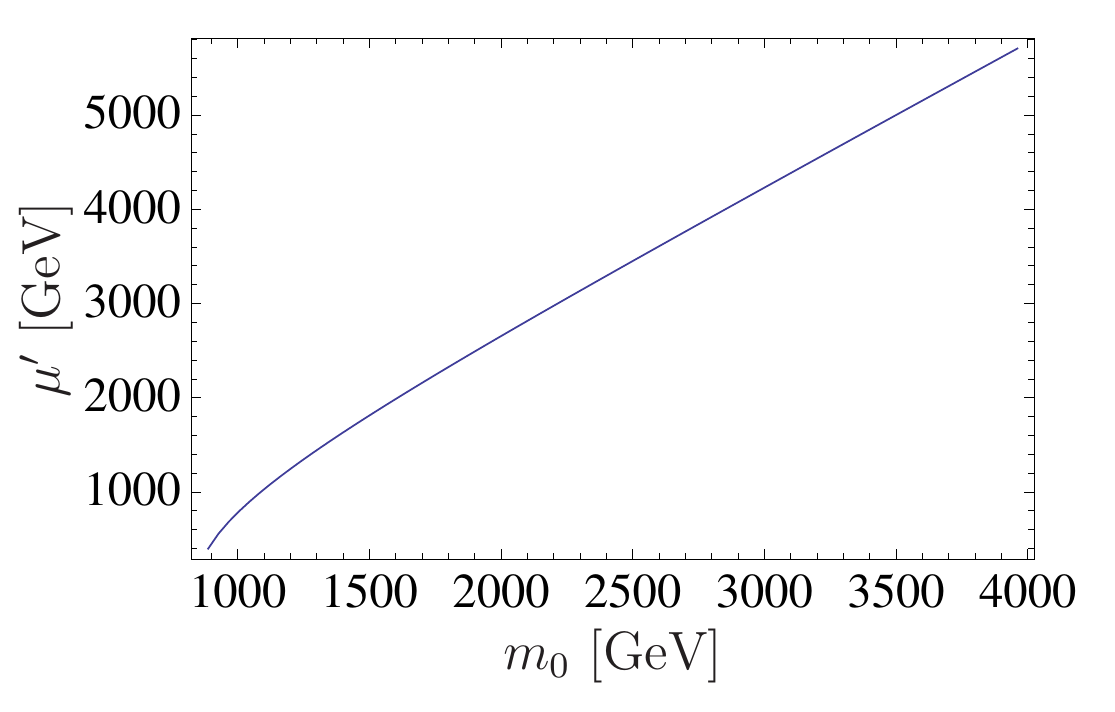} 
\hfill
\includegraphics[width=0.48\linewidth]{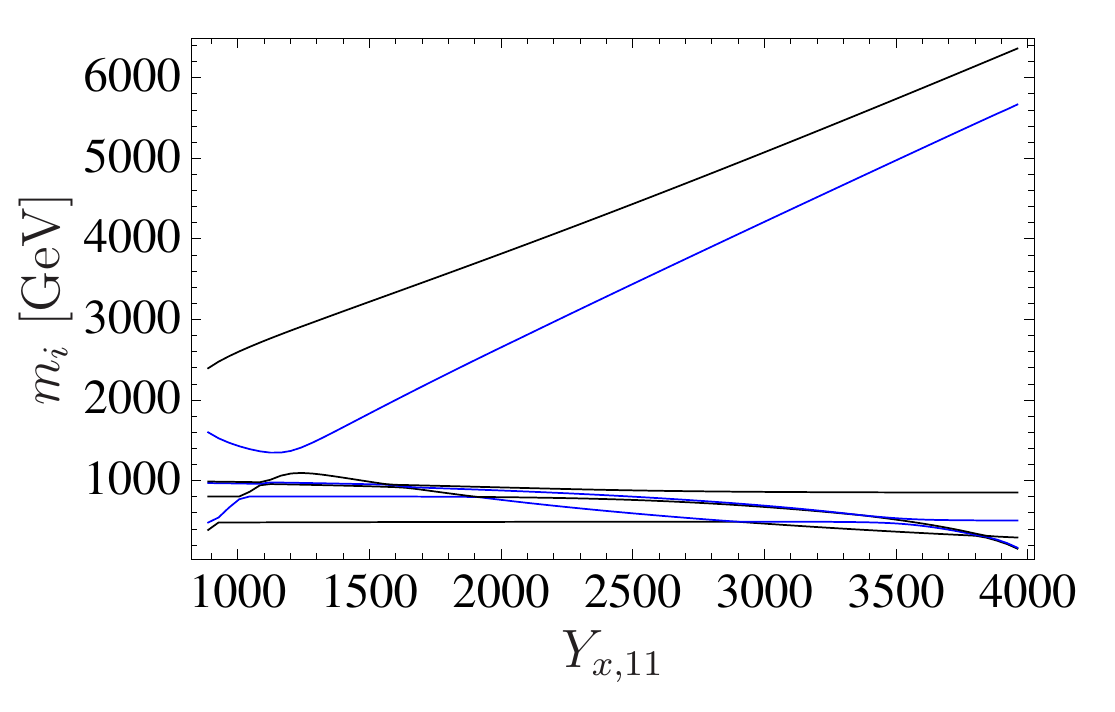} \\ 
 \includegraphics[width=0.48\linewidth]{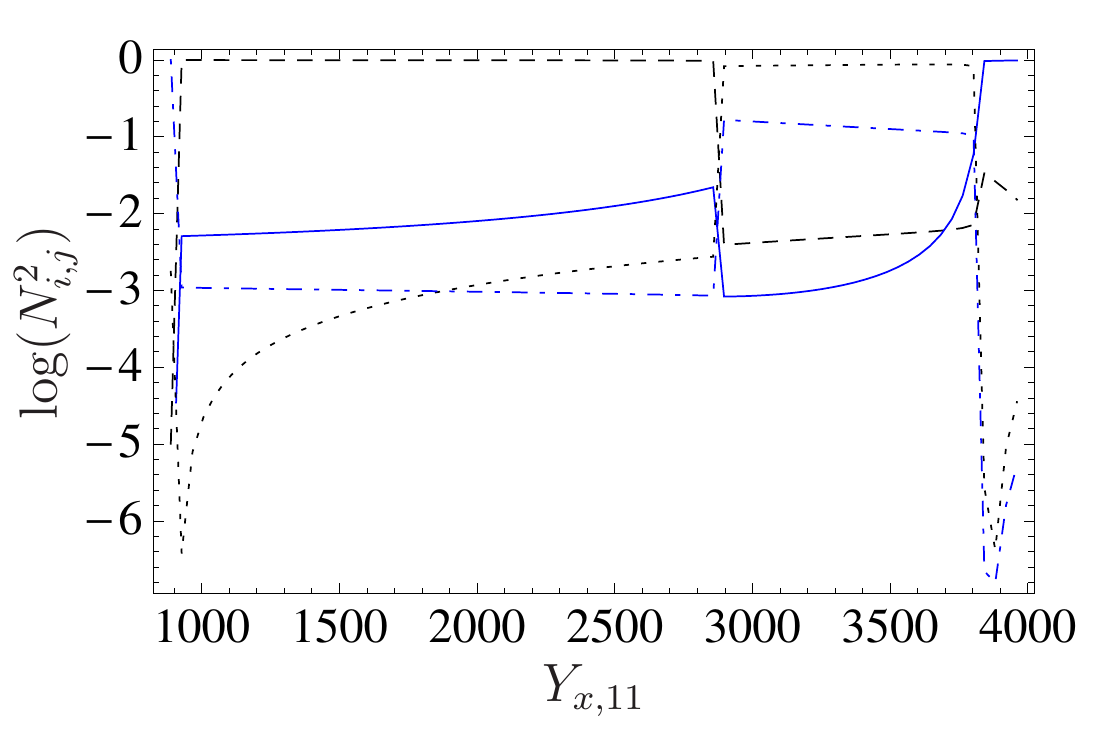} 
   \end{minipage}
\caption{a) $\mu'$ as function of $m_0$. b) masses of all neutralinos.
 c) content of the lightest neutralino: gaugino fraction (dashed black),
 \higgsino
 fraction (blue), logarithm of the \blino fraction (dotted black) and bileptino
 fraction (dot-dashed blue). The
 input parameters were those of \blinoBM but with $M_{1/2} = 1$~TeV.
}
\label{fig:neutralino_BLino}
\end{figure}

As mentioned above, the soft-breaking parameter  $M_{B'}$ is always smaller than
 $M_1$, but the large
mixing between the \blino and the bileptino usually implies that
the bino is still the LSP. However, there are regions
where this mixing is small and the \blino becomes the LSP.
In particular this happens if $\mu' \gg g_{BL} x \simeq M_{Z'}$
which happens either for large $|Y_x|$ or large $m_0$, as this
increases the difference $m_{\bar\eta}^2 - m_\eta^2$.
As an example we show in \FIG~\ref{fig:neutralino_BLino} the dependence
of $\mu'$, the masses of all neutralinos
and the content of the lightest on $m_0$. As claimed,
$\mu'$ grows with increasing $m_0$ leading to a larger mass 
splitting between
the bileptino-like neutralinos and the others. For very large values of
$\mu'$, the bilepton fields are nearly decoupled and
 the nature of the LSP becomes \blino-like. In this case one has
to check if one can obtain the correct value for the relic density.
A principal possibility are resonances as there are two light
Higgs bosons and the LSP mass could easily be half of one of
the Higgs masses. However, it still has to be checked if the 
corresponding couplings are sufficiently large,
which however is beyond the scope of this paper.

\begin{figure}[t]
 \begin{minipage}{0.99\linewidth}
\includegraphics[width=0.48\linewidth]{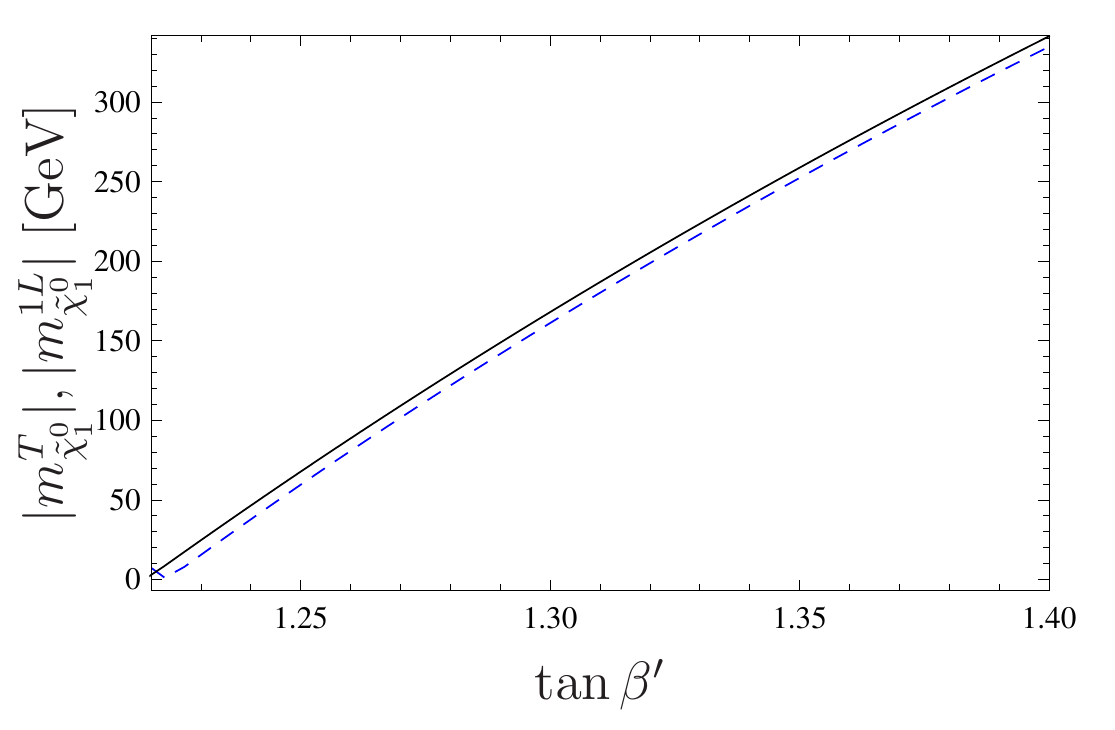} 
\hfill
 \includegraphics[width=0.48\linewidth]{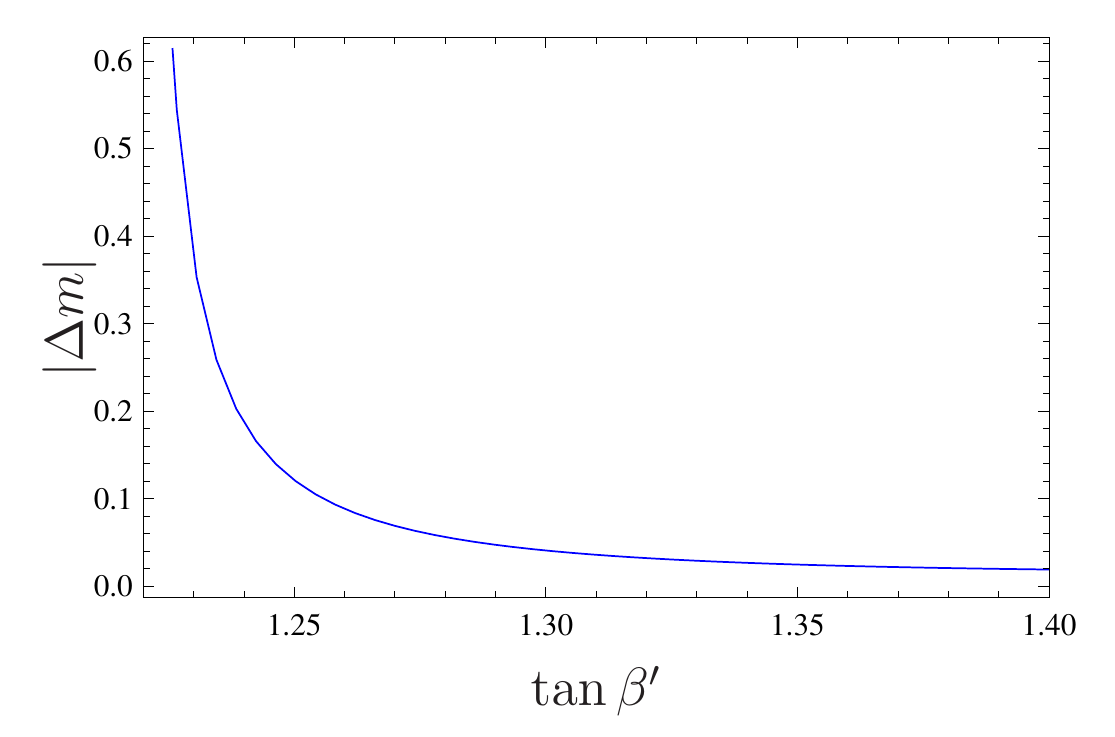} 
   \end{minipage}
\caption{Left: mass of lightest neutralino at tree level (dashed blue) and
one-loop level (black) for a variation of $\tan\beta'$. Right: relative size of
 the correction $|\Delta m| = |1-\frac{m^T}{m^{1L}}|$. The input parameters are
 those of \blinoBM.}
\label{fig:tbp_N1}
\end{figure}

We want to close the discussion of the \blino LSP with a remark about the
importance of the loop corrections. It is well known that in the MSSM one gets a
 few per-cent corrections
to the masses of sleptons, neutralinos and charginos \cite{Pierce:1996zz}. 
In the model considered, the corrections are usually of a similar size.
However, this doesn't apply for a light \blino because here loops contribute with
 rather heavy particles,
in particular $A^0_\eta$ and the bileptino-like neutralinos.
This is demonstrated by inspecting  scenario \blinoBM. 
Varying $\tan\beta'$  we find that these corrections
get larger the closer $\tan\beta'$ gets to one as can be seen in
\FIG~\ref{fig:tbp_N1}. At tree level one
might then conclude that the LSP could be massless even for 
unified gaugino masses.
 This behavior can be roughly understood when neglecting 
 gauge kinetic mixing as then the bileptinos and the BLino decouple
from the MSSM neutralinos. In this limit it is thus
sufficient to consider only the lower
 left $3 \times 3$ block of the neutralino mass matrix given in
 \EQ~(\ref{eq:NeutralinoMM}). Taking $\tan\beta'~1$ or equivalently
$v_{\eta} \simeq v_{\bar{\eta}}$ we find for the LSP mass
\begin{equation}
\label{eq:EV}
\frac{1}{2}(M_B + \mu' - \sqrt{ (M_B - \mu')^2 + 8 (g_{BL} v_{\eta})^2}) 
\end{equation}
This expression can obviously be negative or positive for the same sign of
 $\mu'$ depending on the value of $v_{\eta}$. Although 
this gets changed at the one-loop level, it is still easy to
obtain a dark matter candidate within the mass range preferred by direct detection
 experiments like DAMA \cite{Bernabei:2008yi,Bernabei:2010mq}, 
CRESST \cite{Angloher:2011uu}  or Cogent \cite{Aalseth:2010vx}. 

\begin{figure}[hbt]
 \begin{minipage}{0.99\linewidth}
\includegraphics[width=0.48\linewidth]{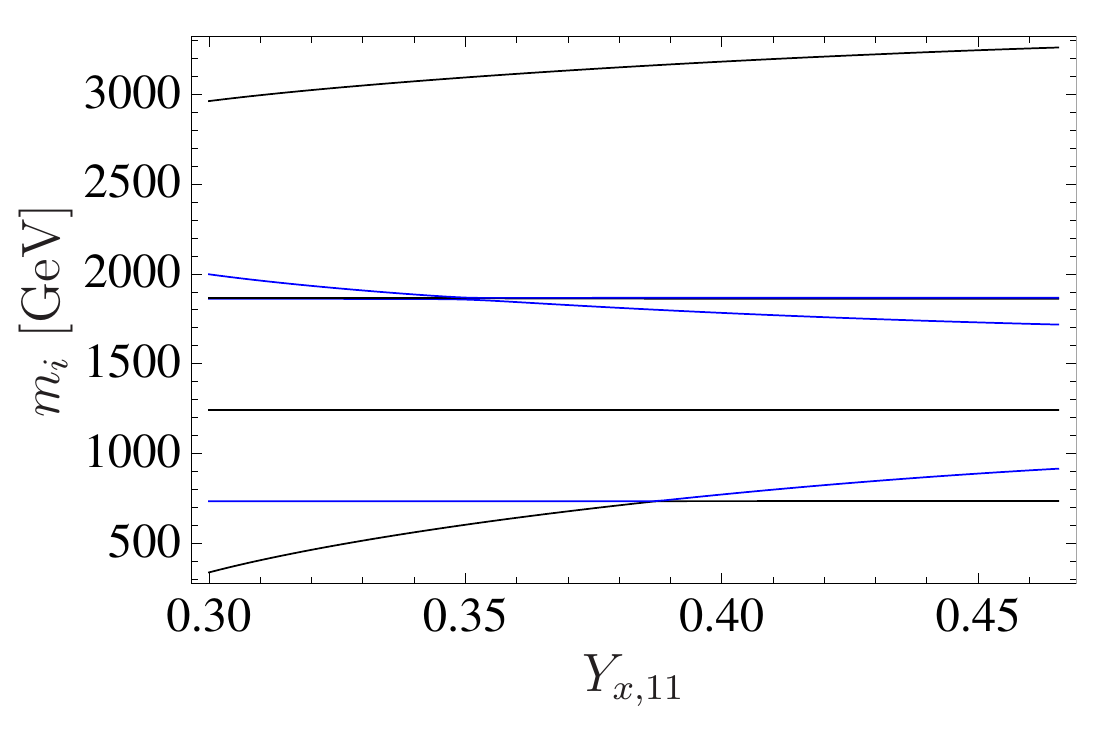}
\hfill
\includegraphics[width=0.48\linewidth]{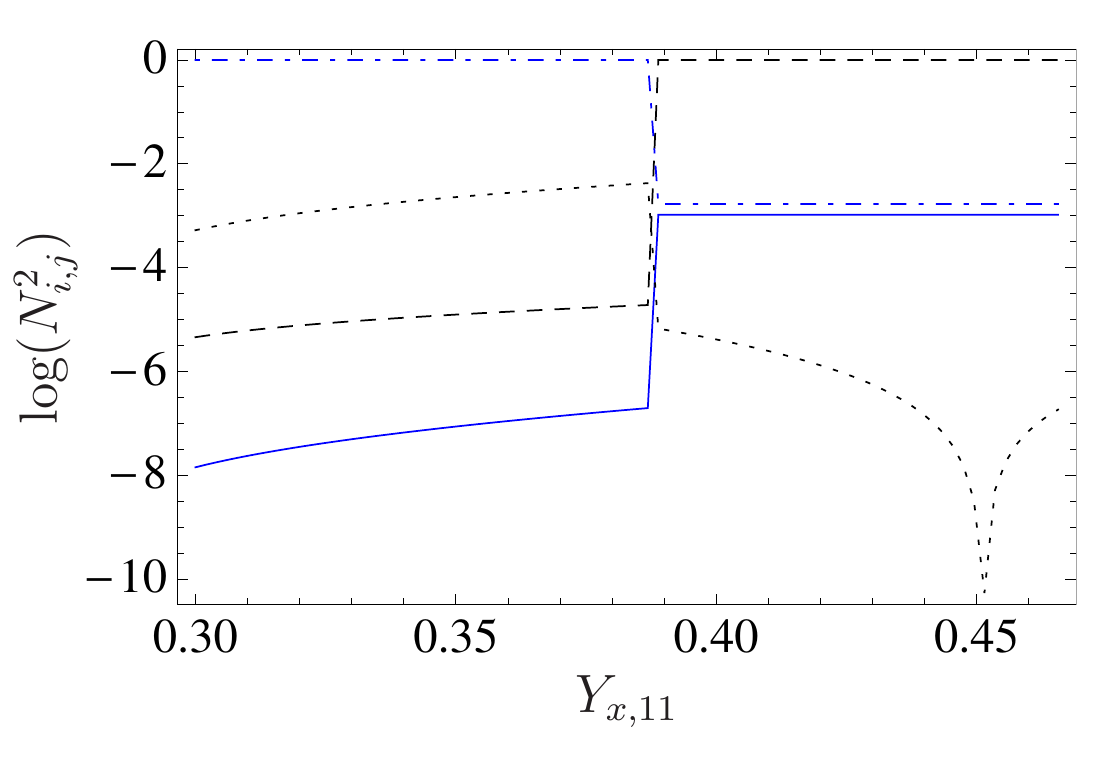}
 \end{minipage}
\caption{LSP with large bileptino fraction (benchmark scenario \bileptinoBM):
 a) mass of
 neutralinos, b) neutralino content. 
The color code on the right-hand side is as follows: gaugino fraction (dashed black),
 \higgsino fraction (blue), logarithm of the \blino fraction (dotted black),
 bileptino fraction (dot-dashed blue).}
\label{fig:Neutralino_Bileptino}
\end{figure}
Finally, we note that also a bileptino-like LSP can be obtained
in this model. The necessary condition, $|\mu'|$ being smaller than 
$|\mu|$ and all
gaugino mass parameters,  can be obtained if the
difference between $m_{\eta}^2$ and
$m_{\bar{\eta}}^2$ becomes small. This can be accommodated by adjusting
the entries of $Y_x$. As an example, we show in
\FIG~\ref{fig:Neutralino_Bileptino} the masses of all neutralinos as
well as the composition of the lightest neutralino as function
of $Y_{x,11}$ while keeping all other values as in 
scenario \bileptinoBM. Already a 10 per-cent decrease leads
to a  nearly a pure bileptino LSP and its mass depends strongly on
$Y_{x,11}$. For larger values a level crossing takes place and the LSP
becomes bino-like. In principal this coupling could be larger
at the electroweak scale but then one would encounter a Landau pole
below the GUT scale.

\section{Conclusions and discussion}
\label{sect:conclusions}
We have discussed in this paper the mass spectrum of the minimal \BL
extension of the MSSM taking universal boundary conditions
at the GUT scale. We have calculated the spectrum using two-loop
RGEs and the complete one-loop contributions to all masses, which
are particularly important in the Higgs and neutralino sectors.
Consistency with current bounds on the
additional $Z'$ implies that the scalar partners of the fermions
are quite heavy in this scenario, except the sneutrinos can be light
 under certain conditions.
 However, this is a consequence
of the taking the mass parameters of the sfermions equal to the ones
in the extended Higgs sector. Relaxing this assumption allows for
lighter non-sneutrino sfermions in addition to light sneutrinos.

It turns out that gauge kinetic mixing between the two 
 Abelian gauge groups is quite important for Higgs bosons
and neutralinos and it cannot be neglected. On one hand it leads to sizable
 shifts in the masses of up to 10 per-cent.
On the other it induces tree-level mixing between
the MSSM states and the states of the extended \BL sector leading
to important shifts in the nature of the corresponding particles.
This holds in particular for the light Higgs bosons and the lightest
neutralino. For example in the latter case we find regions in
parameter space with light neutralinos as preferred by DAMA or
COGENT. Moreover, the nature of the lightest neutralino can be
quite different from the usual CMSSM, \EG we have identified
regions where it is either dominantly \higgsino-,
\blino- or bileptino-like. 

In the extended Higgs sector
we find that one-loop corrections are not only important for
the MSSM-like $h^0$ but also for the light bilepton field. 
This particle can be so light that the MSSM $h^0$-like state can
decay into two of them without conflicting with any of the known
experimental results. However, in general we find that the
corresponding branching ratio is at most a few per-cent.

\section*{Acknowledgements}

We thank Pavel Fileviez Perez, Sogee Spinner, Lorenzo Basso, Stefano Morreti 
and Shaaban Khalil for interesting
discussion. This work has been supported by the
German Ministry of Education and Research (BMBF) under contract 
no.\ 05H09WWEF.

\begin{appendix}

\section{Mass matrices}
\label{app:massmatrices}

Here we collect the tree-level formulas for the remaining sfermion
mass matrices.  
{\allowdisplaybreaks
\begin{itemize}
\item {\bf Mass matrix for Sleptons}, Basis: \( \left(\tilde{e}_{L},
 \tilde{e}_{R}\right) \) 
\begin{equation} 
m^2_{\tilde{e}} = \left( 
\begin{array}{cc}
m_{LL} & \frac{1}{\sqrt{2}} \Big(v_d T_{e}  - v_u \mu^* Y_{e} \Big)\\ 
\frac{1}{\sqrt{2}} \Big(v_d T^\dagger_{e}  - v_u \mu Y^\dagger_{e} \Big)
 & m_{RR}\end{array} 
\right) 
\end{equation} 
\begin{align} 
m_{LL} & =  m_{L}^{2} +\frac{v_d^2}{2} Y^\dagger_{e} Y_{e} + \frac{1}{8} \Big(
 (g_{1}^{2}  -g_{2}^{2} +\tilde{g}^{2}+\tilde{g} g_{BL})(v_{d}^{2}- v_{u}^{2})
 +2(\tilde{g} g_{BL}+ g_{BL}^{2}) (v_{\eta}^{2}  - v_{\bar{\eta}}^{2})\Big)
 {\bf 1}
\label{eq:mLL11}\\ 
m_{RR} & = m_{E}^{2}+\frac{v_d^2}{2} Y_{e} Y^\dagger_{e}
 + \frac{1}{8} \Big((2g_{1}^{2}+2\tilde{g}^{2}+\tilde{g} g_{BL} )
 (v_{u}^{2}- v_{d}^{2})-2\Big(2 \tilde{g} g_{BL} + g_{BL}^{2} \Big)(v_{\eta}^{2}
 - v_{\bar{\eta}}^{2}) \Big)\Big){\bf 1} 
\end{align} 

\item {\bf Mass matrix for Down-Squarks}, Basis:
 \( \left(\tilde{d}_L, \tilde{d}_R\right) \) 
 
\begin{equation} 
m^2_{\tilde{d}} = \left( 
\begin{array}{cc}
m_{LL} & \frac{1}{\sqrt{2}} \Big(v_d T_{d}  - v_u \mu^* Y_{d} \Big)\\ 
\frac{1}{\sqrt{2}} \Big(v_d T^\dagger_{d}  - v_u \mu Y^\dagger_{d} \Big)
 & m_{RR}\end{array} 
\right) 
\end{equation} 
\begin{align} 
m_{LL} &= m_{Q}^{2} +\frac{v_d^2}{2} Y_d^\dagger Y_d  +\frac{1}{24}
 \Big((g_1^2 + 3 g_{2}^{2}+\tilde{g}^{2}+\tilde{g} g_{BL}) (v_{u}^{2}-v_{d}^{2})
+2 (g_{BL}^{2}+\tilde{g} g_{BL})(v_{\bar{\eta}}^{2}-v_{\eta}^{2})\Big)
 {\bf 1} \\
m_{RR} &= m_{D}^{2} +\frac{v_d^2}{2} Y_d Y^\dagger_d+\frac{1}{24}
 \Big(-(2 g_{1}^{2}+2 \tilde{g}^{2}- \tilde{g} g_{BL}) (v_{d}^{2}- v_{u}^{2})
 + 2( g_{BL}^{2}-2\tilde{g} g_{BL} )  (v_{\eta}^{2}- v_{\bar{\eta}}^{2})
 {\bf 1}
\end{align} 

\item {\bf Mass matrix for Up-Squarks}, Basis:
 \( \left(\tilde{u}_L, \tilde{u}_R\right) \) 
\begin{equation} 
m^2_{\tilde{u}} = \left( 
\begin{array}{cc}
m_{LL} &\frac{1}{\sqrt{2}}\Big(v_u T_{u}- v_d \mu^* Y_{u}  \Big)\\ 
\frac{1}{\sqrt{2}} \Big(v_u T^\dagger_{u}- v_d \mu Y^\dagger_{u} \Big)
 & m_{RR}\end{array} 
\right) 
\end{equation} 
\begin{align} 
m_{LL} & = m_{Q}^{2}+\frac{v_u^2}{2} Y^\dagger_u Y_u +\frac{1}{24}
 \Big((g_{1}^{2} -3 g_2^2+\tilde{g}^{2}+\tilde{g} g_{BL} )
 ( v_{u}^{2}- v_{d}^{2}) +2(\tilde{g} g_{BL}+ g_{BL}^{2}) (v_{\bar{\eta}}^{2}
-v_{\eta}^{2})\Big) {\bf 1} \\
m_{RR} & = m_{U}^{2}+\frac{v_u^2}{2} Y_u Y^\dagger_u+ \frac{1}{24}
 \Big(2 (g_{BL}^{2}+4 \tilde{g} g_{BL}) (v_{\eta}^{2}- v_{\bar{\eta}}^{2})
 + (4 g_{1}^{2}+4 \tilde{g}^{2}+ \tilde{g} g_{BL}) ( v_{d}^{2}- v_{u}^{2})\Big)
{\bf 1}
\end{align}

\end{itemize}

\section{RGEs}
\label{app:rges}
The calculation of the renormalization group equations performed by \SARAH is
 based on the generic
expression of  \cite{Martin:1993zk}. In addition, the results of
 \cite{Fonseca:2011vn} are used to
include the effect of kinetic mixing. \\
The $\beta$ functions for the parameters of a general superpotential written as 
\begin{equation}
 W (\phi) = \frac{1}{2}{\mu}^{ij}\phi_i\phi_j + \frac{1}{6}Y^{ijk}
\phi_i\phi_j\phi_k
\end{equation}
can be easily obtained from the shown results for the anomalous dimensions by
 using the relations \cite{West:1984dg,Jones:1984cx}
\begin{eqnarray}
 \beta_Y^{ijk} &= & Y^{p(ij} {\gamma_p}^{k)} \thickspace, \\
 \beta_{\mu}^{ij} &= & \mu^{p(i} {\gamma_p}^{j)} \thickspace .
\end{eqnarray}
For the results of the other parameters as well as for the two-loop results
 which we skip here because of their length we suggest to use
the function {\tt CalcRGEs[]} of \SARAH with the model files shown in
 appendix~\ref{app:modelfiles}.

\subsection{Anomalous dimensions}
\begin{align} 
\gamma_{\hat{q}}^{(1)} & = {Y_{d}^{\dagger}  Y_d} + {Y_{u}^{\dagger}  Y_u}
\nonumber \\
& -\frac{1}{60} \Big(2 (g_{YY}^{2}  + g_{Y B}^{2} )
   + 2 \sqrt{10} ( g_{YY} g_{B Y}+ g_{Y B} g_{B B})
      + 5 \Big(18 g_{2}^{2}  + 32 g_{3}^{2}
  + g_{B B}^{2} + g_{B Y}^{2}\Big)\Big){\bf 1} \\ 
\gamma_{\hat{l}}^{(1)} & =  
 {Y_{e}^{\dagger}  Y_e} + {Y_{\nu}^*  Y_{v}^{T}}\nonumber \\
 &- \frac{3}{20} \Big(2 (g_{YY}^{2} + g_{Y B}^{2})
     + 2 \sqrt{10} (g_{YY} g_{B Y}+  g_{Y B} g_{B B} ) + 
 5 \Big(2 g_{2}^{2}  + g_{B B}^{2} + g_{B Y}^{2}\Big)\Big){\bf 1} \Big)\\ 
\gamma_{\hat{H}_d}^{(1)} & =  
3 \mbox{Tr}\Big({Y_d  Y_{d}^{\dagger}}\Big)
  +  \mbox{Tr}\Big({Y_e  Y_{e}^{\dagger}}\Big)
  -\frac{3}{10} \Big(5 g_{2}^{2}  + g_{YY}^{2} + g_{Y B}^{2}\Big)\\ 
\gamma_{\hat{H}_u}^{(1)} & =  
3 \mbox{Tr}\Big({Y_u  Y_{u}^{\dagger}}\Big)
  +  \mbox{Tr}\Big({Y_{\nu}  Y_{v}^{\dagger}}\Big)
  -\frac{3}{10} \Big(5 g_{2}^{2}  + g_{YY}^{2} + g_{Y B}^{2}\Big)\\ 
\gamma_{\hat{d}}^{(1)} & =  2 Y_d^*  Y_{d}^{T}\nonumber \\
&+ \frac{1}{60} \Big(4 \sqrt{10} (g_{YY} g_{B Y}+ g_{Y B} g_{B B})
    - 5 \Big(32 g_{3}^{2}  + g_{B B}^{2} + g_{B Y}^{2}\Big) 
   - 8 (g_{YY}^{2}  + g_{Y B}^{2} )\Big){\bf 1}\\ 
\gamma_{\hat{u}}^{(1)} & =  2 Y_u^*  Y_{u}^{T} \nonumber \\
& - \frac{1}{60}  \Big(32 (g_{YY}^{2}  + g_{Y B}^{2})
  + 5 \Big(32 g_{3}^{2}  + g_{B B}^{2} + g_{B Y}^{2}\Big) 
 + 8 \sqrt{10} (g_{YY} g_{B Y}  + g_{Y B} g_{B B} )\Big){\bf 1} \\ 
\gamma_{\hat{e}}^{(1)} & =  
2 {Y_e^*  Y_{e}^{T}}  -\frac{3}{20}
 \Big(4 \sqrt{10} (g_{YY} g_{B Y}+ g_{Y B} g_{B B} )
  + 5 \Big(g_{B B}^{2} + g_{B Y}^{2}\Big)
 + 8 (g_{YY}^{2}  +  g_{Y B}^{2}) \Big){\bf 1} \\ 
\gamma_{\hat{\nu}}^{(1)} & =  
2 {Y_{v}^{\dagger}  Y_{\nu}}
  + 2 \Big( Y_{x}^{\dagger}  Y_x  + Y_x^*  Y_x \Big)
  -\frac{3}{4} \Big(g_{B B}^{2} + g_{B Y}^{2}\Big){\bf 1}\\ 
\gamma_{\hat{\eta}}^{(1)} & =  
-3 \Big(g_{B B}^{2} + g_{B Y}^{2}\Big)
 + 2 \mbox{Tr}\Big({Y_x  Y_{x}^{\dagger}}\Big)\\ 
\gamma_{\hat{\bar{\eta}}}^{(1)} & =  
-3 \Big(g_{B B}^{2} + g_{B Y}^{2}\Big)
\end{align} }

\subsection{Gauge Couplings}
We give here and in the subsequent section 
the beta functions for the RGEs of the gauge couplings and gaugino
mass parameters
in a basis independent way as the off-diagonal values for the
$U(1)$ gauge couplings and gaugino mass paramters 
are generated due to RGE effects.
{\allowdisplaybreaks  \begin{align} 
\beta_{g_{YY}}^{(1)} & =  
\frac{3}{5} \Big(11 g_{YY}^{3}  + 4 \sqrt{10} g_{YY}^{2} g_{B Y}
  + g_{YY} \Big(11 g_{Y B}^{2}  + 15 g_{B Y}^{2}
  + 2 \sqrt{10} g_{Y B} g_{B B} \Big) \nonumber \\
& + g_{Y B} \Big(15 g_{B B}  + 2 \sqrt{10} g_{Y B} \Big)g_{B Y} \Big)\\ 
\beta_{g_{B B}}^{(1)} & =  
\frac{3}{5} \Big(11 g_{Y B}^{2} g_{B B} +4 \sqrt{10} g_{Y B} g_{B B}^{2}
 +15 g_{B B}^{3} +11 g_{YY} g_{Y B} g_{B Y} +2 \sqrt{10} g_{YY} g_{B B} g_{B Y}
 \nonumber \\
& +2 \sqrt{10} g_{Y B} g_{B Y}^{2} +15 g_{B B} g_{B Y}^{2} \Big)\\ 
\beta_{g_{Y B}}^{(1)} & =  
\frac{3}{5} \Big(g_{YY} \Big(15 g_{B B}  + 2 \sqrt{10} g_{Y B} \Big)g_{B Y}
  + g_{YY}^{2} \Big(11 g_{Y B}  + 2 \sqrt{10} g_{B B} \Big)  \nonumber \\
& \label{eq:betaGYB} + g_{Y B} \Big(11 g_{Y B}^{2}  + 15 g_{B B}^{2}
  + 4 \sqrt{10} g_{Y B} g_{B B} \Big)\Big)\\ 
\beta_{g_{B Y}}^{(1)} & =  
\frac{3}{5} \Big(11 g_{YY}^{2} g_{B Y}  + g_{YY} \Big(11 g_{Y B} g_{B B}
  + 2 \sqrt{10} \Big(2 g_{B Y}^{2}  + g_{B B}^{2}\Big)\Big)   \nonumber \\
& \label{eq:betaGBY}+
 g_{B Y} \Big(15 \Big(g_{B B}^{2} + g_{B Y}^{2}\Big)
 + 2 \sqrt{10} g_{Y B} g_{B B} \Big)\Big)  \\ 
\beta_{g_2}^{(1)} & =  
g_{2}^{3}\\ 
\beta_{g_3}^{(1)} & =  
-3 g_{3}^{3}
\end{align}}
 
\subsection{Gaugino Mass Parameters}
{\allowdisplaybreaks  \begin{align} 
\beta_{M_{1}}^{(1)} & =  
\frac{6}{5} \Big(11 g_{YY}^{2} M_1 +g_{B Y} \Big(15 g_{B B} {M}_{B B'}
  + 15 g_{B Y} M_1  + 2 \sqrt{10} g_{Y B} {M}_{B B'} \Big)\nonumber \\ 
 &+g_{YY} \Big(11 g_{Y B} {M}_{B B'}  + 2 \sqrt{10} g_{B B} {M}_{B B'}
  + 4 \sqrt{10} g_{B Y} M_1 \Big)\Big)\\ 
\beta_{M_2}^{(1)} & =  
2 g_{2}^{2} M_2 \\ 
\beta_{M_3}^{(1)} & =  
-6 g_{3}^{2} M_3 \\ 
\nonumber \beta_{{M}_{B}}^{(1)} & =  
\frac{6}{5} \Big(11 g_{Y B}^{2} {M}_{B}  + 15 g_{B B} \Big(g_{B B} {M}_{B}
  + g_{B Y} {M}_{B B'} \Big) + \\
& 2 \sqrt{10} g_{Y B} \Big(2 g_{B B} {M}_{B}  + g_{B Y} {M}_{B B'} \Big)
 +   g_{YY} \Big(11 g_{Y B}  + 2 \sqrt{10} g_{B B} \Big){M}_{B B'} \Big)\\ 
\nonumber \beta_{{M}_{B B'}}^{(1)} & =  
\frac{3}{5} \Big(11 g_{YY}^{2} {M}_{B B'} +11 g_{Y B}^{2} {M}_{B B'}
 +  2 \sqrt{10} g_{Y B} \Big(2 g_{B B} {M}_{B B'}
  + g_{B Y} \Big(M_1 + {M}_{B}\Big)\Big)+ \\ 
\nonumber &  15 \Big(g_{B B}^{2} {M}_{B B'}
  + g_{B B} g_{B Y} \Big(M_1 + {M}_{B}\Big) + g_{B Y}^{2} {M}_{B B'} \Big)
\nonumber \\ 
 &+g_{YY} \Big(11 g_{Y B} \Big(M_1 + {M}_{B}\Big)
 + 2 \sqrt{10} \Big(2 g_{B Y} {M}_{B B'}
  + g_{B B} \Big(M_1 + {M}_{B}\Big)\Big)\Big)\Big)
\end{align}}

\section{Model files for \SARAH}
\label{app:modelfiles}
Below we list the  model files used for \SARAH to study the model
presented in this paper. Using this one can generate the
Fortran code for the  \SPheno extension to reproduce the results
presented. These files will
also become part of the public \SARAH package in near future.
 
\subsection{\tt B-L-SSM.m}
\lstset{basicstyle=\scriptsize,
frame=shadowbox}
\begin{lstlisting}
(*-------------------------------------------*)
(*   Particle Content*)
(*-------------------------------------------*)

(* Gauge Superfields *)

Gauge[[1]]={B,   U[1], hypercharge, g1,False};
Gauge[[2]]={WB, SU[2], left,        g2,True};
Gauge[[3]]={G,  SU[3], color,       g3,False};
Gauge[[4]]={Bp,  U[1], BminusL,         g1p, False};

(* Chiral Superfields *)

Fields[[1]] = {{uL,  dL},  3, q,   1/6, 2, 3, 1/6};  
Fields[[2]] = {{vL,  eL},  3, l,  -1/2, 2, 1, -1/2};
Fields[[3]] = {{Hd0, Hdm}, 1, Hd, -1/2, 2, 1, 0};
Fields[[4]] = {{Hup, Hu0}, 1, Hu,  1/2, 2, 1, 0};

Fields[[5]] = {conj[dR], 3, d,  1/3, 1, -3, -1/6};
Fields[[6]] = {conj[uR], 3, u, -2/3, 1, -3, -1/6};
Fields[[7]] = {conj[eR], 3, e,    1, 1,  1, 1/2};
Fields[[8]] = {conj[vR], 3, vR,   0, 1,  1, 1/2};

Fields[[9]]  = {C10, 1, C1, 0, 1, 1, -1};
Fields[[10]] = {C20, 1, C2,  0, 1, 1, 1};


(*------------------------------------------------------*)
(* Superpotential *)
(*------------------------------------------------------*)

SuperPotential = { {{1, Yu},{u,q,Hu}}, {{-1,Yd},{d,q,Hd}},
                   {{-1,Ye},{e,l,Hd}}, {{1,\[Mu]},{Hu,Hd}},
                   {{1,Yv},{l,Hu,vR}}, {{-1,MuP},{C1,C2}},
                   {{1,Yn},{vR,C1,vR}}  };

(*-------------------------------------------*)
(* Integrate Out or Delete Particles         *)
(*-------------------------------------------*)

IntegrateOut={};
DeleteParticles={};

(*----------------------------------------------*)
(*   ROTATIONS                                  *)
(*----------------------------------------------*)

(* ----- Different eigenstates: gauge eigenstates and eigenstates after EWSB
 ---- *)

NameOfStates={GaugeES, EWSB};

(* ----- Gauge fixing terms for Gauge eigenstates ---- *)

DEFINITION[GaugeES][GaugeFixing]=
		{ {Der[VWB],  -1/(2 RXi[W])},
  		  {Der[VG],   -1/(2 RXi[G]) }};



(*--- Rotations in gauge sector ---- *)

DEFINITION[EWSB][GaugeSector] =
{  {{VB,VWB[3],VBp},{VP,VZ,VZp},ZZ},
  {{VWB[1],VWB[2]},{VWm,conj[VWm]},ZW},
  {{fWB[1],fWB[2],fWB[3]},{fWm,fWp,fW0},ZfW}};

       
(*--- VEVs ---- *)

DEFINITION[EWSB][VEVs]= 
{{SHd0, {vd, 1/Sqrt[2]}, {sigmad, \[ImaginaryI]/Sqrt[2]},{phid,1/Sqrt[2]}},
 {SHu0, {vu, 1/Sqrt[2]}, {sigmau, \[ImaginaryI]/Sqrt[2]},{phiu,1/Sqrt[2]}},
 {SvL, {0, 0}, {sigmaL, \[ImaginaryI]/Sqrt[2]},{phiL,1/Sqrt[2]}},
 {SvR, {0, 0}, {sigmaR, \[ImaginaryI]/Sqrt[2]},{phiR,1/Sqrt[2]}},
 {SC10, {x1, 1/Sqrt[2]}, {sigma1, \[ImaginaryI]/Sqrt[2]},{phi1, 1/Sqrt[2]}},
 {SC20, {x2, 1/Sqrt[2]}, {sigma2, \[ImaginaryI]/Sqrt[2]},{phi2, 1/Sqrt[2]}}};
 

 
(*--- Matter Sector ---- *)
 
DEFINITION[EWSB][MatterSector]= 
{    {{SdL, SdR}, {Sd, ZD}},
     {{SuL, SuR}, {Su, ZU}},
     {{SeL, SeR}, {Se, ZE}},
     {{sigmaL,sigmaR}, {SvIm, ZVI}},
     {{phiL,phiR}, {SvRe, ZVR}},
     {{phid, phiu,phi1, phi2}, {hh, ZH}},
     {{sigmad, sigmau,sigma1, sigma2}, {Ah, ZA}},
     {{SHdm,conj[SHup]},{Hpm,ZP}},
     {{fB, fW0, FHd0, FHu0,fBp,FC10,FC20}, {L0, ZN}}, 
     {{{fWm, FHdm}, {fWp, FHup}}, {{Lm,UM}, {Lp,UP}}},
     {{FvL,conj[FvR]},{Fvm,UV}},
     {{{FeL},{conj[FeR]}},{{FEL,ZEL},{FER,ZER}}},
     {{{FdL},{conj[FdR]}},{{FDL,ZDL},{FDR,ZDR}}},
     {{{FuL},{conj[FuR]}},{{FUL,ZUL},{FUR,ZUR}}}                    \
       }; 


(*--- Gauge Fixing after EWSB ---- *)

DEFINITION[EWSB][GaugeFixing]=
  {	{Der[VP],                                            - 1/(2 RXi[P])},	
	{Der[VWm]+\[ImaginaryI] Mass[VWm] RXi[W] Hpm[{1}],   - 1/(RXi[W])},
	{Der[VZ] - Mass[VZ] RXi[Z] Ah[{1}],                  - 1/(2 RXi[Z])},
	{Der[VZp] - Mass[VZp] RXi[Zp] Ah[{2}],               - 1/(2 RXi[Zp])},
	{Der[VG],                                            - 1/(2 RXi[G])}};


(*--- Phases ---- *)

DEFINITION[EWSB][Phases]= 
{    {fG, PhaseGlu}     }; 



(*----------------------------------------------*)
(*   Dirac Spinors                              *)
(*----------------------------------------------*)

(* Dirac Spinors for gauge eigenstates *)

DEFINITION[GaugeES][DiracSpinors]={
  Bino ->{fB, conj[fB]},
  Wino -> {fWB, conj[fWB]},
  Glu -> {fG, conj[fG]},
  H0 -> {FHd0, conj[FHu0]},
  HC -> {FHdm, conj[FHup]},
  Fd1 -> {FdL, 0},
  Fd2 -> {0, FdR},
  Fu1 -> {FuL, 0},
  Fu2 -> {0, FuR},
  Fe1 -> {FeL, 0},
  Fe2 -> {0, FeR},
  Fv1 -> {FvL, 0},
  Fv2 -> {0, FvR},
  FC -> {FC10, conj[FC20]},
  FB -> {fBp, conj[fBp]}
};

(* Dirac Spinors for eigenstates after EWSB *)

DEFINITION[EWSB][DiracSpinors]={
 Fd ->{  FDL, conj[FDR]},
 Fe ->{  FEL, conj[FER]},
 Fu ->{  FUL, conj[FUR]},
 Fv ->{  Fvm, conj[Fvm]},
 Chi ->{ L0, conj[L0]},
 Cha ->{ Lm, conj[Lp]},
 Glu ->{ fG, conj[fG]}
};	
\end{lstlisting}

\subsection{\tt SPheno.m}
\begin{lstlisting}
(*----------------------------------------------*)
(*   MINPAR                                     *)
(*----------------------------------------------*)

MINPAR={{1,m0},
        {2,m12},
        {3,TanBeta},
        {4,SignumMu},
        {5,Azero},
        {6,SignumMuP}, 
        {7,TanBetaP},
        {8,MZp}};

RealParameters = {TanBeta, TanBetaP};

(*----------------------------------------------*)
(*   Tadpoles and renormalization scale         *)
(*----------------------------------------------*)

ParametersToSolveTadpoles = {B[\[Mu]],B[MuP],\[Mu],MuP};

RenormalizationScaleFirstGuess = m0^2 + 4 m12^2;
RenormalizationScale = MSu[1]*MSu[6];


(*----------------------------------------------*)
(*   Boundary conditions                        *)
% (*----------------------------------------------*)

(* ---- Definition of GUT scale ---- *)
ConditionGUTscale = (g1*g1p-g1g1p*g1pg1)/Sqrt[g1p^2+g1pg1^2] == g2;

(* ---- Boundary conditions at GUT scale  ---- *)
BoundaryHighScale={
{g1,(g1*g1p-g1g1p*g1pg1)/Sqrt[g1p^2-g1pg1^2]},
{g1,Sqrt[(g1^2+g2^2)/2]},
{g2,g1},
{g1p, g1},
{g1g1p,0},
{g1pg1,0},
{T[Ye], Azero*Ye},
{T[Yd], Azero*Yd},
{T[Yu], Azero*Yu},
{T[Yv], Azero*Yv},
{T[Yn], Azero*Yn},
{mq2, DIAGONAL m0^2},
{ml2, DIAGONAL m0^2},
{md2, DIAGONAL m0^2},
{mu2, DIAGONAL m0^2},
{me2, DIAGONAL m0^2},
{mvR2, DIAGONAL m0^2},
{mHd2, m0^2},
{mHu2, m0^2},
{mC12, m0^2},
{mC22, m0^2},
{MassB, m12},
{MassWB,m12},
{MassG,m12},
{MassBp,m12},
{MassBBp,0},
{MassBpB,0}
};


(* ---- Boundary conditions at SUSY scale  ---- *)
BoundarySUSYScale = {
 {g1T,(g1*g1p-g1g1p*g1pg1)/Sqrt[g1p^2+g1pg1^2]},
 {g1pT, Sqrt[g1p^2+g1pg1^2]},
 {g1g1pT,(g1g1p*g1p+g1pg1*g1)/Sqrt[g1p^2+g1pg1^2]},
 {g1, g1T},
 {g1p, g1pT},
 {g1g1p, g1g1pT},
 {g1pg1,0},
 {vevP, MZp/g1p},
 {betaP,ArcTan[TanBetaP]},
 {x2,vevP*Cos[betaP]},
 {x1,vevP*Sin[betaP]},
 {Yv,  LHInput[Yv]},
 {Yn,  LHInput[Yn]}
};

(* ---- Boundary conditions at EWSB scale  ---- *)
BoundaryEWSBScale = {
 {g1T,(g1*g1p-g1g1p*g1pg1)/Sqrt[g1p^2+g1pg1^2]},
 {g1pT, Sqrt[g1p^2+g1pg1^2]},
 {g1g1pT,(g1g1p*g1p+g1pg1*g1)/Sqrt[g1p^2+g1pg1^2]},
 {g1, g1T},
 {g1p, g1pT},
 {g1g1p, g1g1pT},
 {g1pg1,0},
 {vevP, MZp/g1p},
 {betaP,ArcTan[TanBetaP]},
 {x2,vevP*Cos[betaP]},
 {x1,vevP*Sin[betaP]}
};

(* ---- Initialization values  ---- *)
InitializationValues = {
 {g1p, 0.5},
 {g1g1p, -0.06},
 {g1pg1, -0.06}
 }

(* ---- Boundary conditions for SUSY scale input  ---- *)
BoundaryLowScaleInput={
 {vd,Sqrt[4 mz2/(g1^2+g2^2)]*Cos[ArcTan[TanBeta]]},
 {vu,Sqrt[4 mz2/(g1^2+g2^2)]*Sin[ArcTan[TanBeta]]}
};

(*----------------------------------------------*)
(*   Two and Three body decays                  *)
(*----------------------------------------------*)

ListDecayParticles = Automatic;
ListDecayParticles3B =Automatic;
\end{lstlisting}

\end{appendix}

\bibliographystyle{h-physrev}

\end{document}